\documentclass[sigconf]{acmart}

\usepackage{algorithm}
\usepackage{algpseudocode}
\usepackage{subcaption}
\usepackage{enumitem}
\usepackage{tabularx}
\usepackage{listings}
\usepackage{multirow}

\setlist[itemize]{leftmargin=*}

\AtBeginDocument{%
  }

\copyrightyear{2026}
\acmYear{2026}
\setcopyright{cc}
\acmConference[SIGIR '26]{Proceedings of the 49th International ACM SIGIR Conference on Research and Development in Information Retrieval}{July 20--24, 2026}{Melbourne, VIC, Australia}
\acmBooktitle{Proceedings of the 49th International ACM SIGIR Conference on Research and Development in Information Retrieval (SIGIR '26), July 20--24, 2026, Melbourne, VIC, Australia}
\acmDOI{10.1145/3805712.3809546}
\acmISBN{979-8-4007-2599-9/2026/07}
\begin{document}

\title{DCGL: Dual-Channel Graph Learning with Large Language Models for Knowledge-Aware Recommendation}

\author{Xinchi Zou}
\orcid{0009-0002-1535-7742}
\affiliation{%
	\institution{School of Computer Science and Technology, Huazhong University of Science and Technology}
	\city{Wuhan}
	\country{China}
}
\email{xinchizou@hust.edu.cn}

\author{Tongzhenzhi Su}
\orcid{0009-0003-0618-1148}
\affiliation{%
	\institution{School of Computer Science and Technology, Huazhong University of Science and Technology}
	\city{Wuhan}
	\country{China}
}
\email{tongzhenzhisu@hust.edu.cn}

\author{Jianjun Li}
\authornote{Corresponding author.}
\orcid{0000-0002-5265-7624}
\affiliation{%
	\institution{School of Computer Science and Technology, Huazhong University of Science and Technology}
	\city{Wuhan}
	\country{China}
}
\email{jianjunli@hust.edu.cn}

\author{Yuan Fu}
\orcid{0009-0000-7087-4975}
\affiliation{%
	\institution{School of Software Engineering, Huazhong University of Science and Technology}
	\city{Wuhan}
	\country{China}
}
\email{yuanfu@hust.edu.cn}

\author{Chang Liu}
\orcid{0009-0007-0189-0964}
\affiliation{%
	\institution{School of Computer Science and Technology, Huazhong University of Science and Technology}
	\city{Wuhan}
	\country{China}
}
\email{diu@hust.edu.cn}

\author{Zhiying Deng}
\orcid{0000-0002-7554-0920}
\affiliation{%
	\institution{Laboratory for Artificial Intelligence and New Forms of Education, Central China Normal University}
	\city{Wuhan}
	\country{China}
}
\email{zhiyingdzy@ccnu.edu.cn}

\author{Zhiwei Shen}
\orcid{0009-0000-6488-4229}
\affiliation{%
	\institution{School of Computer Science and Technology, Huazhong University of Science and Technology}
	\city{Wuhan}
	\country{China}
}
\email{shenzhiwei@hust.edu.cn}

\renewcommand{\shortauthors}{Zou et al.}

\begin{abstract}
  Knowledge Graphs (KGs) have proven highly effective for recommendation systems by capturing latent item relationships, while recent integration of Large Language Models (LLMs) has further enhanced semantic understanding and addressed knowledge sparsity issues.
  Nevertheless, current KG\&LLM-based methods still face three main limitations: 1) inadequate modeling of implicit semantics relationships beyond explicit KG links; 2) suboptimal single-channel fusion of ID and LLM embeddings, which often leads to signal interference and blurred representations; and 3) insufficient consideration of user-item interaction frequency variations in recommendation strategies. To address these challenges, we propose the \textbf{D}ual-\textbf{C}hannel \textbf{G}raph \textbf{L}earning (\textbf{DCGL}) framework, featuring three key innovations: 1) a dual-channel architecture that structurally decoupling rich semantic information from user behavioral patterns, preventing early interference; 2) a multi-level contrastive learning mechanism that enhances robustness against KG noise through intra-view contrast and bridges semantic gaps between channels via inter-view alignment; and 3) a dynamic fusion mechanism that adaptively balances semantic generalization and behavioral specificity based on interaction frequency, resolving the cascading limitation. Extensive experiments on four real-world datasets show that DCGL consistently outperforms state-of-the-art methods, yielding substantial improvements in sparse scenarios while maintaining precision for active users. Our code is available at \url{https://github.com/XinchiZou/DCGL}.
\end{abstract}



\maketitle

\section{Introduction}
\label{sec:introduction}

The rapid proliferation of online platforms has exacerbated the long-standing problem of information overload. Modern recommender systems alleviate this issue by modeling user preferences from historical interactions~\cite{gao2023survey}. Among them, Collaborative Filtering (CF) remains a foundational paradigm, valued for its simplicity and efficacy~\cite{he2017neural,he2020lightgcn,rendle2012bpr}. However, methods relying solely on user-item interactions are inherently susceptible to data sparsity and cold-start problems~\cite{togashi2021alleviating}.  To mitigate these limitations, Knowledge Graphs (KGs) have been integrated as external side information, empowering recommendation models to capture diverse, high-order relational dependencies among items~\cite{wang2019kgat,wang2021learning,yang2022knowledge,yang2023knowledge}. More recently, the semantic understanding and reasoning capabilities of Large Language Models (LLMs) have been leveraged to infer user interests~\cite{ren2024representation,zhao2024breaking} and integrate them into collaborative learning~\cite{hu2025bridging}, as well as to comprehend local and global KG structures to compensate for missing facts and sparsity~\cite{cui2025comprehending}.

\begin{figure}[t]
	\centering
	\includegraphics[width=\linewidth]{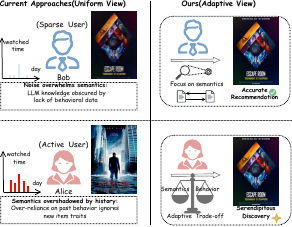}
	\caption{Motivating Example. Addressing frequency heterogeneity by shifting from a uniform view (Left) to an adaptive view (Right) that balances semantic and behavioral signals for different user groups.}
	\label{fig:motivation}
\end{figure}

Despite encouraging progress, we argue that state-of-the-art knowledge-aware recommenders~\cite{cui2025comprehending,ren2024representation} are trapped in a vicious cycle rooted in frequency heterogeneity. User and item interaction frequencies exhibit a highly skewed distribution, yet mainstream methods forcibly employ uniform signal fusion within single-channel architectures. This practice blurs the fundamental distinction between frequency-sensitive behavioral embeddings (ID embeddings) and frequency-agnostic semantic embeddings (LLM embeddings). As a result, entities with rich interactions suffer from semantic dilution, while those with sparse interactions lack sufficient behavioral grounding. This suboptimal fusion subsequently undermines the model's ability to capture implicit semantics—subtle relationships not explicitly encoded in the knowledge graph, such as the affinity between ``mystery'' and ``detective'' genres. Consequently, a chain reaction ensues: frequency skew complicates the fusion process, while poor fusion in turn hinders semantic discovery.

\textbf{Motivating Example}. Consider recommending a new mystery movie ``Escape Room''  to two users: Alice (a frequent viewer with hundreds of interactions) and Bob (a new user with few clicks) on a streaming platform, as shown in Figure~\ref{fig:motivation}. State-of-the-art models that fuse LLM semantics with collaborative IDs in a single channel struggle in both scenarios. For Bob, weak behavioral signals are overwhelmed by noise, drowning out clear semantic cues and leading to inaccurate recommendations. For Alice, strong historical patterns overshadow subtle semantic affinities (e.g., between the ``mind-bending narrative'' in her watched sci‑fi movies and the new mystery film), leading to missed serendipitous discoveries. This reveals a cascading limitation: the inability to adapt fusion to frequency heterogeneity (Bob vs. Alice) results in blurred representations, which in turn hinders the effective use of implicit semantics for robust recommendations across the sparsity spectrum.

To break this cycle, we propose \textbf{D}ual-\textbf{C}hannel \textbf{G}raph \textbf{L}earning (\textbf{DCGL}), a unified KG-enhanced recommendation framework designed to resolve the cascading problem through synergistic mechanisms. Unlike approaches that treat each challenge in isolation, DCGL employs a coherent strategy built on three pillars: (1) \textbf{A dual-channel architecture} that provides the structural basis by decoupling the learning of LLM semantics and ID collaborative signals into parallel pathways, thus preventing early signal interference.(2) \textbf{Multi-level contrastive learning} applied within this structure to explicitly shape robust and aligned representations; intra-view contrast enhances robustness within each channel, while inter-view contrast bridges the semantic and behavioral spaces, enabling the ID channel to internalize implicit semantic nuances. (3) \textbf{A frequency-aware gated fusion} mechanism that dynamically orchestrates the final integration. Crucially, this gating is not independent. Instead, it is informed by the clarified representations from the contrastive-learning-enhanced channels and is modulated by the interaction frequency, thereby enabling adaptive reliance on semantics or behavior according to sparsity levels. Together, these components work in concert: dual channels enable separation, contrastive learning refines and aligns within and across them, and the adaptive gate achieves intelligent, context-sensitive unification.

Our experiments on four real-world datasets show that DCGL consistently outperforms strong baselines. Ablation studies confirm the contribution of each component and their interdependencies, and detailed analyses demonstrate substantial gains in sparse and cold-start scenarios. Our main contributions are threefold:
\begin{itemize}
	\item We identify and formalize a cascading limitation in knowledge-aware recommendation, where frequency heterogeneity, suboptimal single-channel fusion, and inadequate modeling of implicit semantics exacerbate one another.
	\item We propose DCGL, a unified framework that tackles this core problem through the synergistic design of a dual-channel architecture, multi-level contrastive learning, and a frequency-aware gated fusion mechanism.
	\item We provide comprehensive validation showing DCGL resolves the trade-off between semantic generalization and behavioral specificity, boosting sparse and long-tail performance while preserving precision for active users.
\end{itemize}

\begin{figure*}[t]
	\centering
	\includegraphics[width=0.99\linewidth]{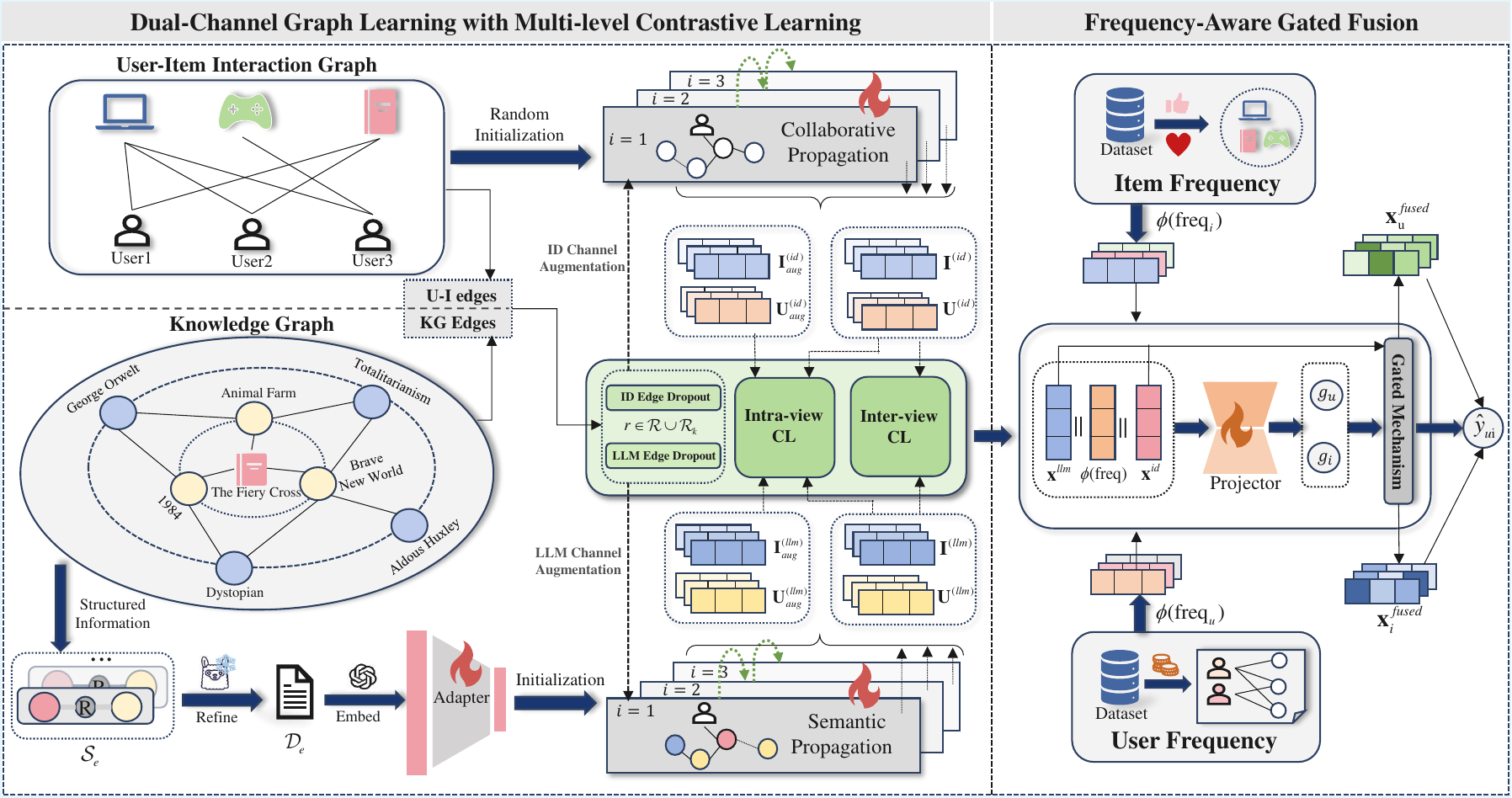}
	\caption{The framework of our proposed DCGL.}
	\label{fig:model}
\end{figure*}

\section{Related work}
\label{sec:relatedwork}

\subsection{Knowledge-aware Recommendation}
Knowledge-aware recommendation is a key research area within recommender systems, encompassing embedding-based, path-based, and graph neural network (GNN)-based methodologies in recent advancements. Embedding-based approaches~\cite{zhang2016collaborative,wang2018dkn,cao2019unifying} leverage relations and entities from knowledge graphs to enrich user and item representations. 

Path-based methods~\cite{yu2014personalized,Wang_Wang_Xu_He_Cao_Chua_2019}, on the other hand, focus on extracting meta-paths from knowledge graphs to capture user-item connectivity patterns. 

Progressing further, GNN-based methods~\cite{wang2019kgat,wang2021learning,yang2022knowledge,yang2023knowledge,zou2022multi,zou2024knowledge,li2025hypercomplex,li2025lightkg}, at the research forefront, refine entity and relation representations by aggregating multi-hop neighbor embeddings. Notably, KGAT~\cite{wang2019kgat} employs graph attention mechanisms to propagate embeddings based on neighbor importance, enabling end-to-end recommendation score generation;
KGRec~\cite{yang2023knowledge} evaluates triplets with rationale scores, combining MAE-based reconstruction learning with contrastive objectives to improve recommendation performance. However, reliance on incomplete, explicit KG paths leaves these models brittle to missing facts and weak at capturing implicit semantics.

\subsection{LLMs for Recommendation}
In recent years, applying LLMs to recommender systems has emerged as a research hotspot~\cite{zhao2024recommender,chen2024large,wu2024survey}, with LLMs primarily serving two roles: as recommenders or as enhancers. When acting as recommenders, LLMs integrate recommendation tasks into conversational contexts by tokenizing user and item IDs to directly generate recommendations~\cite{li2024large}. Benefiting from their extensive world knowledge, LLMs perform well in zero-shot~\cite{hou2024large} and few-shot scenarios~\cite{bao2023tallrec}, while also offering benefits in fairness~\cite{gao2025sprec} and explainability~\cite{li2025g}. 
However, when used as recommenders, LLMs face challenges such as high inference costs, poor top-$K$ calibration, and underuse of collaborative filtering, hindering their scalable deployment.
As enhancers, LLMs generate user-item interactions, item features~\cite{wei2024llmrec}, or semantic embeddings~\cite{ren2024representation,jiang2025reclm} to augment traditional recommendation models. However, current enhancer methods typically fuse ID embeddings with LLM-generated semantic embeddings in a single-channel manner, failing to distinguish their unique roles in capturing behavioral signals and semantic knowledge. This limits the effective integration of LLMs’ rich semantic information with the behavioral signals captured by ID embeddings, thereby constraining the accurate modeling of complex user preferences.

\subsection{KG\&LLM-based Recommendation}
Recently, hybrid approaches combining KGs and LLMs have gained popularity for alleviating knowledge sparsity and enriching semantic understanding. For example, CIKGRec~\cite{hu2025bridging} uses LLMs to infer user interests, builds a Collaborative Interest KG, and enhances user representations via interest reconstruction and cross-domain contrastive learning. CoLaKG~\cite{cui2025comprehending} fills missing KG facts by extracting subtrees and using prompt engineering for local information, while incorporating global knowledge through semantic retrieval, optimizing recommendations via representation fusion and retrieval-augmented learning.
Despite their potential, these methods still have some limitations: CIKGRec mainly enriches user-side semantics but ignores item-side ones, while CoLaKG  heavily depends on LLM accuracy and retrieval quality, risking noise or hallucinations with large or noisy KGs. Moreover, both use static strategies to integrate semantic and collaborative signals, lacking dynamic adaptation to data sparsity or preference changes.

\section{Preliminaries}
\label{sec:preliminaries}

\noindent\textbf{User-Item Interaction Graph.}
Let $\mathcal{U}$ and $\mathcal{I}$ denote the user set and item set in the recommender system, respectively. We construct a bipartite user-item graph $\mathcal{G}=(\mathcal{U},\mathcal{I},\mathcal{R})$, where $\mathcal{R}=\{(u,i)\mid u\in \mathcal{U},i\in \mathcal{I},{{y}_{ui}}=1\}$ represents the set of interactions between users and items, with ${{y}_{ui}}=1$ indicating that user $u$ has interacted with item $i$, and ${{y}_{ui}}=0$ otherwise. The neighbor sets of user $u$ and item $i$ are defined as ${{\mathcal{N}}_{u}}=\{i\in \mathcal{I}\mid (u,i)\in \mathcal{R}\}$ and ${{\mathcal{N}}_{i}}=\{u\in \mathcal{U}\mid (u,i)\in \mathcal{R}\}$, respectively.

\vspace{0.5em}
\noindent\textbf{Knowledge Graph.}
We incorporate real-world knowledge via a heterogeneous graph $\mathcal{G}_{k}=\{(h,r,t)\}$, defined over an entity set $\mathcal{E}$ and a relation set $\mathcal{R}_k$. Here, $(h,r,t)$ denotes a triplet where $h,t \in \mathcal{E}$ are entities and $r \in \mathcal{R}_k$ is a relation. Items in the interaction graph are linked to corresponding entities in $\mathcal{E}$. Formally, we define the item set $\mathcal{I}$ as a subset of the entity set $\mathcal{E}$ ($\mathcal{I} \subseteq \mathcal{E}$), which enables the modeling of rich relational dependencies. For an item $i\in \mathcal{I}$, its adjacent entities in $\mathcal{G}_{k}$ are given by ${{\mathcal{N}}_{i}^{k}}=\{e\in \mathcal{E}\mid (i,r,e)\in \mathcal{G}_{k}\}$.

\vspace{0.5em}
\noindent\textbf{Task Definition.}
Given the user-item interaction graph $\mathcal{G}$ and knowledge graph ${{\mathcal{G}}_{k}}$, our goal is to learn a recommendation model that predicts the preference score ${{\hat{y}}_{ui}}$ of user $u$ for item $i$. The model ranks items based on these scores and recommends the top-$K$ items with the highest scores. The learning is performed by using the observed interactions in $\mathcal{R}$ to infer users’ potential preferences.

\section{Methodology}
\label{sec:method}

We now present the detailed technical design of our proposed DCGL. The overall framework is illustrated in Figure~\ref{fig:model}, comprising two main components: (1) Dual-Channel Graph Learning with Multi-level Contrastive Learning (left), where the ID channel captures behavioral patterns through collaborative propagation while the LLM channel models semantic knowledge via semantic propagation, with multi-level contrastive learning enhancing both robustness and cross-channel alignment; (2) Frequency-Aware Gated Fusion (right), which adaptively integrates dual-channel representations based on interaction frequencies to produce the final prediction.

\subsection{Dual-Channel Graph Learning Architecture}

To break the vicious cycle of signal interference in single-channel designs, we introduce a dual-channel graph learning architecture. This structure provides the foundational separation needed to decouple the learning of frequency-sensitive behavioral patterns (ID channel) from frequency-agnostic semantic knowledge (LLM channel), preventing the semantic dilution and grounding issues illustrated in our motivating example.

\subsubsection{\textbf{Entity Embedding Generation}}

\begin{figure}[t]
	\centering
	\includegraphics[width=\linewidth]{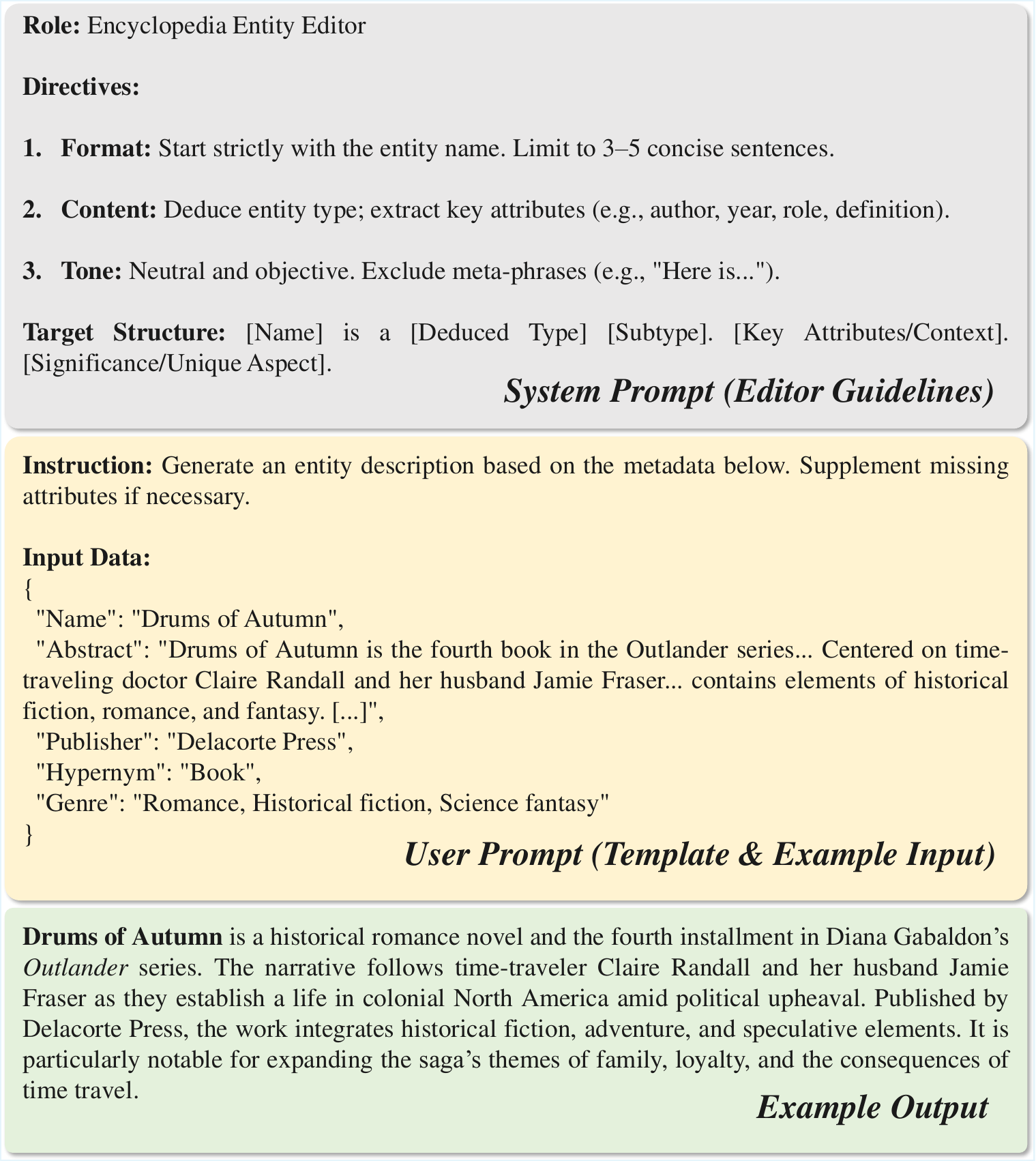}
	\caption{Prompt template for LLM-based entity description, including system prompt with editing rules, user prompt example and generated output.}
	\label{fig:llm-prompt-template}
\end{figure}

In line with our goal of distinct signal separation, we generate entity embeddings through two parallel, specialized pipelines.

\vspace{0.5em}

\noindent \textbf{LLM Semantic Channel:} To capture rich, transferable semantics, we employ an entity-level description-to-embedding pipeline using Large Language Models (LLMs). As illustrated in Figure~\ref{fig:llm-prompt-template}, structured fields of each entity $e$, denoted as $\mathcal{S}_e$ (e.g., name, attributes), are injected into a reusable template. We guide the LLM to synthesize these attributes into a coherent, neutral-toned textual description $\mathcal{D}_{e}$, effectively capturing nuanced semantic relationships beyond superficial metadata. This description is then encoded by a pre-trained text embedding model to derive a high-dimensional semantic representation:
\begin{equation}
\mathbf{x}_{e}^{(\text{llm,raw})} = \text{TextEmbedding}(\mathcal{D}_{e}) \in \mathbb{R}^{d_{\text{llm}}}
\end{equation}
A trainable adapter then projects this high-dimensional vector into the model's standard embedding space $\mathbb{R}^{d}$:
\begin{equation}
	\mathbf{x}_e^{(\text{llm})} = \mathbf{W}_2 \cdot \text{LeakyReLU}\left( \mathbf{W}_1 \mathbf{x}_e^{(\text{llm,raw})} + \mathbf{b}_1 \right) + \mathbf{b}_2,
\end{equation}
where $\mathbf{W}_1, \mathbf{W}_2, \mathbf{b}_1, \mathbf{b}_2$ are learnable parameters. This process yields semantic embeddings that are independent of interaction frequency.

\vspace{0.5em}
\noindent \textbf{ID Information Channel:} To preserve precise collaborative filtering signals from historical interactions, embeddings in this channel $\mathbf{x}_{e}^{(\text{id})} \in \mathbb{R}^{d}$ are randomly initialized and optimized purely on user-item interaction data, creating a dedicated pathway for learning behavioral patterns that are highly sensitive to interaction frequency.

This explicit separation at the input stage establishes the clear distinction between semantic and behavioral signals that is crucial for our subsequent adaptive fusion.

\subsubsection{\textbf{Knowledge Graph and Interaction Signal Modeling}}
We model relational information and collaborative effects within each separate channel to maintain signal purity.

\vspace{0.5em}
\noindent \textbf{Relation-Aware Knowledge Graph Encoding:} 
Inspired by recent advancements in knowledge graph-based methods~\cite{wang2019kgat,yang2023knowledge}, we employ a relation-aware graph attention network (RGAT)
in each channel $c \in \{\text{id}, \text{llm}\}$ to encode the knowledge graph $\mathcal{G}_k$. This approach dynamically models entity-relation dependencies through parameterized attention mechanisms, thereby obviating the need for manual path specification. Specifically, for an item $i$ with relational neighbors $\mathcal{N}_i^{k}$, the RGAT layer performs:
\begin{equation}
	\begin{gathered}
		\mathbf{x}_{i}^{(c,l+1)} = \mathbf{x}_{i}^{(c,l)} + \sum\nolimits_{e \in \mathcal{N}_{i}^{k}} a_{e,i}^{(c,l)} \mathbf{x}_{e}^{(c,l)}, \\
		a_{e,i}^{(c,l)} = \frac{\exp \left( \phi (\mathbf{r}_{e,i}^{(c,l)}, \mathbf{x}_{e}^{(c,l)}, \mathbf{x}_{i}^{(c,l)}) \right)}{\sum_{e' \in \mathcal{N}_{i}^{k}} \exp \left( \phi (\mathbf{r}_{e',i}^{(c,l)}, \mathbf{x}_{e'}^{(c,l)}, \mathbf{x}_{i}^{(c,l)}) \right)}.
	\end{gathered}
\end{equation}
where the attention coefficient $\phi$ is defined by a LeakyReLU-activated bilinear projection:
\begin{equation}
	\phi ({{\mathbf{r}_{e,i}}},{{\mathbf{x}}_{e}},{{\mathbf{x}}_{i}})=\text{LeakyReLU}\left( {{{{\mathbf{r}_{e,i}}}}^{\top }}\mathbf{W}[{{\mathbf{x}}_{e}}\|{{\mathbf{x}}_{i}}] \right),
\end{equation}
with $\mathbf{W} \in \mathbb{R}^{d \times 2d}$ denoting the learnable projection matrix, and $\mathbf{r}_{e,i}^{(c)} \in \mathbb{R}^d$ representing the relation-specific embedding in channel $c$. The relation embeddings $\mathbf{r}_{e,i}^{(c)}$ in each channel are independently optimized: the ID channel learns behavior-aware relation representations from user-item interaction data, while the LLM channel incorporates semantic knowledge to enhance relational modeling. After $L$ layers of RGAT aggregation, the item representation matrix $\mathbf{I}^{(c)} = \{ \mathbf{x}_{i}^{(c,L)} \mid i \in \mathcal{I} \}$ encodes both semantic information and relational structures from $\mathcal{G}_k$.

\vspace{0.5em}
\noindent \textbf{Collaborative Signal Propagation:}
To further model the collaborative effects in user-item interactions, we employ a graph-based collaborative filtering framework to encode user and item representations. Specifically, we adopt LightGCN~\cite{he2020lightgcn}'s efficient message propagation strategy to capture high-order collaborative signals through $L'$ propagation layers:
\begin{align}
	\mathbf{x}_{u}^{(c,l+1)} &= \sum\nolimits_{i \in \mathcal{N}_{u}} \frac{\mathbf{x}_{i}^{(c,l)}}{\sqrt{|\mathcal{N}_{u}| \cdot |\mathcal{N}_{i}|}}, \\
	\mathbf{x}_{i}^{(c,l+1)} &= \sum\nolimits_{u \in \mathcal{N}_{i}} \frac{\mathbf{x}_{u}^{(c,l)}}{\sqrt{|\mathcal{N}_{i}| \cdot |\mathcal{N}_{u}|}},
\end{align}
where $\mathbf{x}_{u}^{(c,l)}$ and $\mathbf{x}_{i}^{(c,l)}$ represent the embeddings of user $u$ and item $i$ at the $l$-th graph propagation layer in channel $c$, respectively. Note here $\mathbf{x}_{u}^{(c,0)}$ is randomly initialized, while $\mathbf{x}_{i}^{(c,0)}$ is initialized with RGAT output $\mathbf{x}_{i}^{(c,L)}$. After stacking $L'$ LightGCN layers, each channel $c$ yields the user representation matrix $\mathbf{U}^{(c)}=\{\mathbf{x}_{u}^{(c)}\mid u\!\in\!\mathcal{U}\}\!\in\!\mathbb{R}^{|\mathcal{U}|\times d}$
and the item representation matrix $\mathbf{I}^{(c)}=\{\mathbf{x}_{i}^{(c)}\mid i\!\in\!\mathcal{I}\}\!\in\!\mathbb{R}^{|\mathcal{I}|\times d}$.
The LLM channel thus provides semantically-oriented embeddings,
whereas the ID channel focuses on behavioral signals.

\subsubsection{\textbf{Semantic Regularization via Translation-based Training.}}
To further enhance the geometric consistency of the entity and relation embeddings used in RGAT, we incorporate an auxiliary translation-based training objective~\cite{bordes2013translating}. We adopt an alternating optimization strategy between the graph propagation and the TransE objective in each channel $c \in \{\text{id}, \text{llm}\}$. The core idea is to enforce the structure $\mathbf{h}^{(c)} + \mathbf{r}^{(c)} \approx \mathbf{t}^{(c)}$, where $\mathbf{h}, \mathbf{r}$ and $\mathbf{t}$ correspond to the embeddings of the head entity, relation, and tail entity in a triplet $(h,r,t) \in \mathcal{G}_k$, respectively. The distance score is defined using the $L_1$ norm:
\begin{equation}
	f_d^{(c)}(h, r, t) = \|\mathbf{h}^{(c)} + \mathbf{r}^{(c)} - \mathbf{t}^{(c)}\|_1.
\end{equation}
We employ a pairwise ranking loss to discriminate valid triplets from corrupted ones $(h, r, t')$, which are generated via negative sampling by randomly replacing the tail entity. The regularization loss for channel $c$ is then formulated as:
\begin{equation}
	\mathcal{L}_{\text{TransE}}^{(c)} = \sum_{(h,r,t,t') \in \mathcal{G}_k} -\ln \sigma \big( f_d^{(c)}(h, r, t') - f_d^{(c)}(h, r, t) \big),
\end{equation}
where $\sigma(\cdot)$ is the sigmoid function. This objective optimizes the embeddings such that valid triplets have significantly smaller distances than corrupted ones. Specifically, the loss in the ID channel refines behavior-aware structural embeddings, while in the LLM channel, it ensures that the semantic embeddings maintain valid relational dependencies. By alternating the optimization of $\mathcal{L}_{\text{TransE}}^{(c)}$ with the main recommendation task, DCGL effectively regularizes the representation space without interfering with the gradient flow of the collaborative filtering process.
\subsection{Multi-level Contrastive Learning}
After generating the semantically and behaviorally dominant representations $\mathbf{U}^{(c)}$ and $\mathbf{I}^{(c)}$, we introduce a multi-level contrastive learning mechanism to enhance the model's robustness against noise, sparsity, and cross-channel heterogeneity. Here, we treat each representation variant as a distinct \textit{view} of the same entity—graph augmentation produces augmented views within each channel, while the dual-channel outputs naturally form behavioral and semantic views. This mechanism comprises two complementary components: intra-view augmentation for structural stability preservation and inter-view alignment for cross-channel representation calibration.

\subsubsection{\textbf{Intra-View Augmentation}}
Building upon graph augmentation strategies~\cite{yang2022knowledge}, we implement a two-stage procedure that first perturbs the knowledge graph $\mathcal{G}_k$ and then performs stability-aware perturbation on the user--item interaction graph $\mathcal{G}$. For $\mathcal{G}_k$, we generate augmented views through random edge dropout:
\begin{equation}
	\mathcal{G}_k' \sim \text{DropEdge}(\mathcal{G}_k, \rho),
\end{equation}
where $\rho \in (0,1)$ denotes the edge dropout rate. Both original and augmented graphs are encoded by channel-specific RGAT encoders, yielding item-level stability scores $S_i \in [0,1]$ via KG edge consistency measurement as defined in~\cite{yang2022knowledge}. Conditioned on these stability scores, we perform stability-guided interaction graph augmentation:
\begin{equation}
	\begin{gathered}
		\mathcal{G}' \sim \text{StabAdaptDrop}(\mathcal{G}, p_{\text{drop}}), \\
		p_{\text{drop}}(u,i) = 1 - \mu S_i,
	\end{gathered}
\end{equation}
where $(u,i)$ represents an interaction edge and $\mu \in (0,1]$ controls the stability guidance strength (i.e., higher $S_i$ leads to lower drop probability).

The intra-view contrastive objective is formulated as the sum of losses over both channels:
\begin{equation}
	\mathcal{L}_{\text{aug}}
	= \sum_{c \in \{\text{id},\text{llm}\}}
	\Big[\mathcal{L}_{\text{InfoNCE}}(\mathbf{U}_{\text{aug}}^{(c)},\mathbf{U}^{(c)})
	+ \mathcal{L}_{\text{InfoNCE}}(\mathbf{I}_{\text{aug}}^{(c)},\mathbf{I}^{(c)})\Big],
\end{equation}
where $\mathbf{U}^{(c)} \in \mathbb{R}^{|\mathcal{U}| \times d}$ and $\mathbf{I}^{(c)} \in \mathbb{R}^{|\mathcal{I}| \times d}$ denote the original embeddings, and $\mathbf{U}_{\text{aug}}^{(c)}, \mathbf{I}_{\text{aug}}^{(c)}$ represent their augmented counterparts. The InfoNCE loss is defined using cosine similarity:
\begin{equation}
	\mathcal{L}_{\text{InfoNCE}}(\mathbf{Z}_1,\mathbf{Z}_2)
	= -\sum_{n=1}^{N}
	\log \frac{\exp(\text{sim}(\mathbf{z}_{1,n},\mathbf{z}_{2,n})/\tau)}
	{\sum_{m \neq n} \exp(\text{sim}(\mathbf{z}_{1,n},\mathbf{z}_{2,m})/\tau)},
	\label{eq:infonce}
\end{equation}
where $\mathbf{Z}_1, \mathbf{Z}_2 \in \mathbb{R}^{N \times d}$ are embedding matrices from two distinct views, $\text{sim}(\cdot,\cdot)$ denotes cosine similarity, $\tau > 0$ is a temperature hyperparameter, and $N$ corresponds to the batch size (i.e., $|\mathcal{U}|$ or $|\mathcal{I}|$). This design enhances representation robustness by preserving stable structural information across both $\mathcal{G}_k$ and $\mathcal{G}$, thereby increasing resilience to noisy or sparse interactions.

\subsubsection{\textbf{Inter-View Alignment}}
To bridge the inherent heterogeneity between the ID and LLM representation spaces, we introduce an inter-view alignment mechanism. This module employs lightweight projection networks to map user embeddings from both channels into a unified latent space. Structurally, each projection network consists of a linear layer that preserves the embedding dimension, followed by a LeakyReLU activation to introduce nonlinearity. Formally, the projection is defined as:
\begin{equation}
	\mathbf{U}_{\text{proj}}^{(\text{id})} = \text{Proj}_{1}(\mathbf{U}^{(\text{id})}), \quad \mathbf{U}_{\text{proj}}^{(\text{llm})} = \text{Proj}_{2}(\mathbf{U}^{(\text{llm})}).
\end{equation}
We optimize the alignment via an InfoNCE-based objective, consistent with the intra-view contrastive loss:
\begin{equation}
	\mathcal{L}_{\text{align}} = \mathcal{L}_{\text{InfoNCE}}(\mathbf{U}_{\text{proj}}^{(\text{id})}, \mathbf{U}_{\text{proj}}^{(\text{llm})}),
\end{equation}
where $\mathcal{L}_{\text{InfoNCE}}$ follows the formulation in Eq.\eqref{eq:infonce}. Specifically, embeddings of the same user from the projected ID and LLM spaces constitute positive pairs, whereas pairings with different users serve as negatives. Maximizing the mutual information between these views not only reduces the semantic gap but also forces the ID channel to implicitly internalize semantic nuances (e.g., genre affinities) from the LLM channel. This ensures that the behavioral representations remain semantically aware even when the ID channel dominates the final fusion.

\subsection{Frequency-Aware Gated Fusion}
To adaptively integrate dual-channel representations, we propose a frequency-aware gating mechanism that dynamically modulates the contributions of the ID channel and the LLM channel according to item popularity. This design enables context-sensitive fusion by balancing behavioral patterns (captured by the ID channel) and semantic knowledge (provided by the LLM channel) through popularity-informed gating.

\subsubsection{\textbf{Gating Network Design}}
The gating network computes fusion weights by incorporating three complementary signals: ID representations, LLM representations, and interaction frequency statistics. For any item $i \in \mathcal{I}$, the gating weight is formulated as:
\begin{equation}
	g_i = \sigma\left(\mathbf{W}_g \left[\mathbf{x}_i^{(\text{id})} \| \mathbf{x}_i^{(\text{llm})} \| \phi(\text{freq}_i)\right] + \mathbf{b}_g\right),
\end{equation}
where $\phi(\text{freq}_i) = \log(1 + \text{freq}_i)/\log(1 + \max_{j \in \mathcal{I}} \text{freq}_j)$ denotes the logarithmically normalized interaction frequency, $\text{freq}_i = |\mathcal{N}_i|$ is the number of interactions involving item $i$, $\mathbf{W}_g \in \mathbb{R}^{1 \times (2d + 1)}$ and $\mathbf{b}_g \in \mathbb{R}$ are learnable parameters, and $\sigma$ is the sigmoid activation.  The user-side gating weight $g_u$ is defined analogously, using user-specific features and interaction statistics.

\subsubsection{\textbf{Adaptive Fusion}}
Based on the derived gating weights, we produce fused representations through a weighted concatenation:
\begin{equation}
	\label{eq:fused_embedding}
	\mathbf{x}_{i}^{\text{fused}}=[{{g}_{i}}\cdot \mathbf{x}_{i}^{(\text{id})}\|(1-{{g}_{i}})\cdot \mathbf{x}_{i}^{(\text{llm})}],
\end{equation}
with an identical formulation for user representations $\mathbf{x}_{u}^{\text{fused}}$. This fusion strategy achieves three key objectives: 
\begin{itemize}[leftmargin=2em]
	\item \textbf{High-frequency regime:} 
	it emphasises the ID channel  ($\uparrow g_i$) to leverage precise  behavioral patterns from abundant interaction data.
	\item \textbf{Low-frequency regime:} it shifts weight toward the LLM channel  ($\downarrow g_i$) to mitigate data sparsity issues by relying on external knowledge.
	\item \textbf{Interpretability:} the explicit gating weights offer insights for model diagnosis, as visualized in Section~\ref{sec:gating_vis}.
\end{itemize}

To compute the preference score, we first consider the direct inner product of the fused representations:
\begin{equation}
	\begin{aligned}
		\text{score}_{ui}^{\text{raw}} &= (\mathbf{x}_{u}^{\text{fused}})^{\top} \mathbf{x}_{i}^{\text{fused}} \\
		&= g_u g_i \underbrace{(\mathbf{x}_u^{(\text{id})})^{\top} \mathbf{x}_i^{(\text{id})}}_{s_{\text{id}}} + (1-g_u)(1-g_i) \underbrace{(\mathbf{x}_u^{(\text{llm})})^{\top} \mathbf{x}_i^{(\text{llm})}}_{s_{\text{llm}}}.
	\end{aligned}
\end{equation}
However, the total weight mass $g_u g_i + (1-g_u)(1-g_i)$ can vary substantially with the gating values, which may cause scale imbalance and destabilise training. To address this, we introduce a normalization factor $\alpha_{ui} = g_u g_i + (1 - g_u)(1 - g_i)$. The final preference score is then defined as:
\begin{equation}
	\hat{y}_{ui} = \frac{(\mathbf{x}_{u}^{\text{fused}})^{\top} \mathbf{x}_{i}^{\text{fused}}}{\alpha_{ui}} = \frac{g_u g_i s_{\text{id}} + (1-g_u)(1-g_i) s_{\text{llm}}}{g_u g_i + (1-g_u)(1-g_i)}.
\end{equation}
Letting $w = \frac{g_u g_i}{\alpha_{ui}}$, the score can be rewritten as a convex combination $\hat{y}_{ui} = w \cdot s_{\text{id}} + (1-w) \cdot s_{\text{llm}}$ with $w \in (0,1)$. This formulation ensures that the combined score remains on the same numerical scale as the individual channel scores regardless of the absolute values of $g_u$ and $g_i$, thereby eliminating scale drift and maintaining  smooth gradient flow. Finally, the top-$K$ items with the highest $\hat{y}_{ui}$ scores are selected as recommendations.

\subsubsection{\textbf{Gating Regularization}}
To prevent channel dominance and ensure balanced fusion, we add a gating regularisation term that penalises extreme weight distributions. For each training triple $(u, i, i') \in \mathcal{O}$ (user $u$, positive item $i$, and negative item $i'$), the regularization loss is:
\begin{equation}
	\mathcal{L}_{\text{gate}} = \frac{1}{|\mathcal{O}|} \sum_{(u, i, i') \in \mathcal{O}} \sum_{x \in \{u, i, i'\}} \left[ -\log g_x - \log(1 - g_x) \right],
\end{equation}
where ${{g}_{x}}\in (0,1)$ denotes the scalar gating weight. This loss maximizes the log-likelihood of $g_x$ and $1 - g_x$, promoting a uniform weight distribution in the interval $(0,1)$ and thus enhancing fusion stability and recommendation performance.

\begin{table}[t]
	\centering
	\small
	\caption{Statistics of the four evaluation datasets.}
	\vspace{-1em}
	\label{tab:dataset_stats}
	\setlength{\tabcolsep}{2.5pt}
	\begin{tabular}{lrrrr}
		\toprule
		\textbf{Statistics} & \textbf{Book-Crossing} & \textbf{DBbook} & \textbf{MovieLens} & \textbf{Amazon-Book} \\
		\midrule
		\#Users     & 6,616   & 5,576   & 6,040   & 11,356 \\
		\#Items     & 8,853   & 2,680   & 3,260   & 9,736 \\
		\#Ratings   & 110,662 & 65,961  & 998,539 & 303,633 \\
		\#Density   & 0.19\%  & 0.44\%  & 5.07\%  & 0.27\% \\
		\midrule
		\#Relations & 4       & 13      & 20      & 10 \\
		\#Entities  & 1,404   & 8,762   & 14,377  & 2,707 \\
		\#Triplets  & 1,137   & 134,223 & 415,104 & 91,886 \\
		\bottomrule
	\end{tabular}
\end{table}

\subsection{Model Training}
The DCGL model is trained in a supervised manner by minimising a composite loss function that integrates the dynamic weighted BPR loss, intra‑view contrastive loss, inter‑view alignment loss, gating regularisation, and $L_2$ regularization:
\begin{equation}
	\mathcal{L}={{\mathcal{L}}_{\text{BPR}}}+{{\lambda }_{\text{aug}}}{{\mathcal{L}}_{\text{aug}}}+{{\lambda }_{\text{align}}}{{\mathcal{L}}_{\text{align}}}+{{\lambda }_{\text{gate}}}{{\mathcal{L}}_{\text{gate}}}+{{\lambda }_{\text{reg}}}\|\boldsymbol{\Theta} \|_{2}^{2}.
\end{equation}
Where $\boldsymbol{\Theta}$ denotes all trainable parameters, and ${{\lambda }_{\text{aug}}}$, ${{\lambda }_{\text{align}}}$, ${{\lambda }_{\text{gate}}}$, ${{\lambda }_{\text{reg}}}$ are weighting coefficients. The dynamic weighted BPR loss is defined as:
\begin{equation}
	\mathcal{L}_{\text{BPR}} = \sum_{(u,i,i') \in \mathcal{O}} -\alpha_{ui} \log \sigma \left( \hat{y}_{ui} - \hat{y}_{ui'} \right),
\end{equation}
where $\mathcal{O} = \{(u, i, i') \mid (u, i) \in \mathcal{R}, (u, i') \notin \mathcal{R}\}$ is the set of training triples, $\hat{y}_{ui}$ is the  preference score, $\alpha_{ui}$ serves as a normalisation factor that stabilises the score scale,  and $\sigma$ is the sigmoid function. 

\subsection{Complexity Analysis}
We analyze the time complexity of DCGL from three key components. (1) \textbf{Dual‑channel graph propagation:}  The RGAT module in the KG channel involves attention calculations with complexity $O(L|\mathcal{R}_k|d^2)$, while the LightGCN module in the interaction channel takes $O(L|\mathcal{R}|d)$, where $|\mathcal{R}_k|$ and $|\mathcal{R}|$ denote the number of triplets and interactions, $L$ is the layer count, and $d$ is the embedding dimension. (2) \textbf{Multi‑level contrastive learning:} This component incurs $O(|\mathcal{R}| B d)$ per epoch for calculating pairwise similarities within batch size $B$. (3) \textbf{Frequency‑aware gating:} The gating operations are element‑wise and incur $O((|\mathcal{U}| + |\mathcal{I}|)d)$. Consequently, the overall complexity grows linearly with the scale of the graphs.  Compared with existing LLM‑enhanced methods such as CIKGRec and CoLaKG, DCGL does not inflate the graph size nor perform computationally expensive online retrieval. Because the LLM-based semantic extraction is performed offline, DCGL maintains high training efficiency comparable to state-of-the-art GNN-based knowledge-aware recommendation models.

\begin{table*}[t]
	\centering
	\small
	\caption{Performance comparison of different methods. The best results are \textbf{bolded} with * indicating statistical significance ($p < 0.05$), and the second best results are {underlined}.}
	\label{tab:performance}	
	\setlength{\tabcolsep}{1.5pt}
	\begin{tabular}{lcccccccccccccccc}
		\toprule
		\multirow{2.5}{*}{\textbf{Methods}} & \multicolumn{4}{c}{\textbf{Book-Crossing}} & \multicolumn{4}{c}{\textbf{DBbook}} & \multicolumn{4}{c}{\textbf{MovieLens}} & \multicolumn{4}{c}{\textbf{Amazon-Book}} \\
		\cmidrule(lr){2-5} \cmidrule(lr){6-9} \cmidrule(lr){10-13} \cmidrule(lr){14-17}
		& \textbf{R@50} & \textbf{R@100} & \textbf{N@50} & \textbf{N@100} & \textbf{R@50} & \textbf{R@100} & \textbf{N@50} & \textbf{N@100} & \textbf{R@50} & \textbf{R@100} & \textbf{N@50} & \textbf{N@100} & \textbf{R@50} & \textbf{R@100} & \textbf{N@50} & \textbf{N@100} \\
		\midrule
		LightGCN & 0.1607 & 0.2246 & 0.0765 & 0.0911 & 0.4214 & 0.5358 & 0.2230 & 0.2481 & 0.4086 & 0.5574 & 0.3977 & 0.4432 & 0.2982 & 0.4083 & 0.1573 & 0.1864 \\
		SGL      & 0.1613 & 0.2154 & 0.0860 & 0.0982 & 0.4258 & 0.5249 & 0.2350 & 0.2567 & 0.4094 & 0.5505 & 0.4004 & 0.4433 & 0.2866 & 0.3769 & 0.1583 & 0.1825 \\
		SimGCL   & 0.1653 & 0.2211 & 0.0860 & 0.0988 & 0.4280 & 0.5369 & 0.2376 & 0.2610 & 0.4131 & 0.5560 & 0.4048 & 0.4482 & 0.3099 & 0.4181 & 0.1651 & 0.1938 \\
		\midrule
		CKE      & 0.1284 & 0.1691 & 0.0708 & 0.0811 & 0.3419 & 0.4448 & 0.1832 & 0.2060 & 0.3948 & 0.5421 & 0.3841 & 0.4295 & 0.2834 & 0.3844 & 0.1541 & 0.1808 \\
		CFKG     & 0.1216 & 0.1771 & 0.0503 & 0.0617 & 0.3586 & 0.4559 & 0.1627 & 0.1816 & 0.3930 & 0.5439 & 0.3672 & 0.4142 & 0.2826 & 0.3941 & 0.1312 & 0.1669 \\
		KGAT     & 0.1564 & 0.2133 & 0.0865 & 0.1004 & 0.4174 & 0.5309 & 0.2110 & 0.2360 & 0.4059 & 0.5527 & 0.3928 & 0.4378 & 0.3045 & 0.4144 & 0.1595 & 0.1888 \\
		KGIN     & 0.1468 & 0.2075 & 0.0681 & 0.0819 & 0.4147 & 0.5337 & 0.2200 & 0.2462 & 0.4095 & 0.5581 & 0.4004 & 0.4453 & 0.3324 & 0.4478 & 0.1815 & 0.2123 \\
		KGCL     & 0.1562 & 0.2049 & 0.0859 & 0.0971 & 0.4308 & 0.5271 & 0.2453 & 0.2666 & 0.4083 & 0.5496 & 0.3996 & 0.4430 & 0.3340 & 0.4411 & 0.1854 & 0.2141 \\
		MCCLK    & 0.1476 & 0.2039 & 0.0656 & 0.0774 & 0.4376 & 0.5491 & 0.2054 & 0.2273 & 0.4176 & 0.5655 & 0.3876 & 0.4339 & 0.3285 & 0.4412 & 0.1614 & 0.1890 \\
		KGRec    & 0.1473 & 0.2075 & 0.0699 & 0.0836 & 0.4415 & 0.5497 & 0.2373 & 0.2611 & 0.4136 & 0.5633 & 0.4072 & 0.4524 & 0.3263 & 0.4429 & 0.1787 & 0.2096 \\
		\midrule
		RLMRec   & 0.1682 & 0.2308 & 0.0820 & 0.0963 & 0.4286 & 0.5380 & 0.2375 & 0.2615 & 0.4162 & 0.5618 & 0.4016 & 0.4452 & 0.3161 & 0.4320 & 0.1686 & 0.1991 \\
		CoLaKG   & \underline{0.1888} & \underline{0.2512} & \underline{0.0898} & \underline{0.1044} & 0.4494 & 0.5617 & 0.2400 & 0.2637 & 0.4232 & 0.5692 & 0.4062 & 0.4492 & \underline{0.3427} & \underline{0.4582} & 0.1849 & 0.2156 \\
		CIKGRec  & 0.1791 & 0.2455 & 0.0889 & 0.1041 & \underline{0.4642} & \underline{0.5756} & \underline{0.2494} & \underline{0.2739} & \underline{0.4250} & \underline{0.5718} & \underline{0.4162} & \underline{0.4609} & 0.3391 & 0.4492 & \underline{0.1863} & \underline{0.2157} \\
		\midrule
		\textbf{DCGL} & \textbf{0.1923*} & \textbf{0.2605*} & \textbf{0.0979*} & \textbf{0.1119*} & \textbf{0.4664*} & \textbf{0.5804*} & \textbf{0.2572*} & \textbf{0.2798*} & \textbf{0.4252*} & \textbf{0.5727*} & \textbf{0.4176*} & \textbf{0.4616*} & \textbf{0.3574*} & \textbf{0.4663*} & \textbf{0.2028*} & \textbf{0.2317*} \\
		\bottomrule
	\end{tabular}
\end{table*}

\section{Experiment}
\label{sec:experiment}

\subsection{Experimental Settings}
\subsubsection{\textbf{Datasets}}
We conducted extensive experiments on four real world recommendation datasets: DBbook~\cite{cao2019unifying}, Book‑Crossing~\cite{dong2017hybrid}, MovieLens~\cite{noia2016sprank}, and Amazon‑Book~\cite{he2016ups}. All datasets contain auxiliary knowledge on the item side and cover diverse domains, scales, and densities. Following common practice~\cite{he2020lightgcn,xie2022contrastive}, we filtered out low‑frequency users (users with fewer than 10 interactions in MovieLens and Amazon‑Book, and fewer than 5 interactions in DBbook and Book‑Crossing). The statistics of the processed datasets are summarised in Table~\ref{tab:dataset_stats}.

\subsubsection{\textbf{Baseline}}
For a comprehensive evaluation, we compare DCGL with a wide range of baselines, grouped into five categories: (1) \textbf{Traditional CF:} LightGCN, SGL, SimGCL; (2) \textbf{Embedding based knowledge-aware}: CKE, CFKG; (3) \textbf{GNN-based knowledge-aware:} KGAT, KGIN, KGCL, MCCL, KGRec; (4) \textbf{LLM-enhanced GNN:} RLMRec; and (5) \textbf{LLM-enhanced KG:} CoLaKG, CIKGRec.

\begin{itemize}[leftmargin=2em]

\item \textbf{LightGCN}~\cite{he2020lightgcn} simplifies GCN by removing feature transformation and nonlinear activation,  retaining only linear neighbourhood aggregation.

\item \textbf{SGL}~\cite{wu2021self}  introduces self‑supervised learning by creating contrastive views through node/edge dropout on user–item graph. 

\item \textbf{SimGCL}~\cite{yu2022graph} discards graph‑structure augmentations and instead adds uniform noise to embeddings to generate contrastive views. 

\item \textbf{CKE}~\cite{zhang2016collaborative} integrates collaborative filtering with structural, textual and visual knowledge embeddings via TransR and auto‑encoders.

\item \textbf{CFKG}~\cite{ai2018learning} formulates recommendation as link prediction on a unified graph that includes users, items and entities. 

\item \textbf{KGAT}~\cite{wang2019kgat}  employs graph attention networks to recursively propagate embeddings, with attention weights derived from relation semantics. 

\item \textbf{KGIN}~\cite{wang2021learning} disentangles user intents and aggregates relational paths to capture fine‑grained preferences.. 

\item \textbf{KGCL}~\cite{yang2022knowledge} adopts knowledge‑graph augmentation and cross view contrastive learning to alleviate noise and sparsity. 

\item \textbf{MCCL}~\cite{zou2022multi} proposes multi‑level contrastive learning across collaborative, semantic and structural views.

\item \textbf{KGRec}~\cite{yang2023knowledge} is a self-supervised method that rationalises knowledge connections by masking and reconstruction. 

\item \textbf{RLMRec}~\cite{ren2024representation} enriches item profiles with LLM‑generated textual signals and aligns them with behavioural relations. 

\item \textbf{CoLaKG}~\cite{cui2025comprehending} uses LLMs to comprehend local sub‑graphs and retrieve global neighbours, fusing them to supplement missing KG facts.

\item \textbf{CIKGRec}~\cite{hu2025bridging} infers user interests with LLMs and constructs a Collaborative Interest Knowledge Graph, employing reconstruction modules to reduce noise.
\end{itemize}

\subsubsection{\textbf{Implementation Details and Evaluation Metrics}}
For each user, the historical interactions were randomly split into training, validation and test sets following a 7:1:2 ratio. For DCGL, hyper‑parameters were tuned as follows: learning rate in $\{0.0001,\\0.0002,\dots,0.005\}$; number of RGAT layers $L$ and LightGCN propagation depth $L'$ both set to $3$; embedding dimension fixed to $64$ (consistent with all baselines); gating regularisation weight $\lambda_{\text{gate}}$, intra‑view contrastive weight $\lambda_{\text{aug}}$ and inter‑view contrastive weight $\lambda_{\text{align}}$ searched over $\{0.01,\dots,0.1\}$ with step $0.01$; KG‑edge dropout rate $\rho$ for intra‑view augmentation set to $0.5$; stability‑aware dropout coefficient $\mu$ tuned in $\{0.1,\dots,1.0\}$ with step $0.1$.

For the LLM semantic channel, we first extracted structured information $\mathcal{S}_e$ (e.g., item names and attributes) from DBpedia\footnote{\url{https://dbpedia.org/sparql}}. These structured entries were then fed to Meta‑Llama‑3‑8B‑Instruct\footnote{\url{https://huggingface.co/meta-llama/Meta-Llama-3-8B-Instruct}} to generate textual descriptions. To minimize randomness, we set temperature $\tau=0$ and top‑$p=0.001$. The generated texts were encoded into semantic embeddings using text‑embedding‑3‑small\footnote{\url{https://api.openai.com/v1/embeddings}}. The adapter’s intermediate dimension was set to ${{d}_{\text{mid}}}  = (d_{\text{llm}}+d)/2$, where $d_{\text{llm}}=1536$ and $d=64$. All experiments ran on an NVIDIA GeForce RTX 4090 GPU.  

For all baselines, we used the authors’ official implementations and tuned hyper‑parameters according to their recommendations or validation performance. To ensure fairness, common settings were standardised: embedding dimension $64$, evaluation interval $1$, and early stopping with a maximum of $2,000$ epochs. Reported results are the mean and standard deviation over $10$ independent runs with different random seeds.

We adopt Recall@$K$ (R@$K$) and NDCG@$K$ (N@$K$), $K\in\{50,100\}$, to evaluate Top‑$K$ recommendation accuracy and ranking quality, respectively. The same evaluation protocol is applied to all LLM‑ and text‑embedding‑based baselines.

\begin{table}[t]
	\centering
	\caption{Ablation study of DCGL components on four datasets. R@100 and N@100 are reported.}
	\vspace{-1em}
	\label{tab:ablation}
	\footnotesize
	\setlength{\tabcolsep}{1.5pt}
	\begin{tabular}{lcccccccc}
		\toprule
		\multirow{2.5}{*}{\textbf{Variants}} & \multicolumn{2}{c}{\textbf{Book-Crossing}} & \multicolumn{2}{c}{\textbf{DBbook}} & \multicolumn{2}{c}{\textbf{MovieLens}} & \multicolumn{2}{c}{\textbf{Amazon-Book}} \\
		\cmidrule(lr){2-3} \cmidrule(lr){4-5} \cmidrule(lr){6-7} \cmidrule(lr){8-9}
		& \textbf{R@100} & \textbf{N@100} & \textbf{R@100} & \textbf{N@100} & \textbf{R@100} & \textbf{N@100} & \textbf{R@100} & \textbf{N@100} \\
		\midrule
		\textit{w/o LLM}   & 0.2319 & 0.0985 & 0.5557 & 0.2655 & 0.5565 & 0.4444 & 0.4440 & 0.2145 \\
		\textit{w/o ID}    & 0.2500 & 0.1017 & 0.5573 & 0.2599 & 0.5663 & 0.4495 & 0.4466 & 0.2154 \\
		\textit{CAT}       & 0.2396 & 0.1045 & 0.5619 & 0.2697 & 0.5536 & 0.4458 & 0.4513 & 0.2263 \\
		\midrule
		\textit{w/o FREQ}  & 0.2576 & 0.1116 & 0.5757 & 0.2745 & 0.5687 & 0.4588 & 0.4654 & 0.2308 \\
		\midrule
		\textit{w/o AUG}   & 0.2425 & 0.1010 & 0.5778 & 0.2690 & 0.5706 & 0.4580 & 0.4496 & 0.2121 \\
		\textit{w/o ALIGN} & 0.2578 & 0.1100 & 0.5741 & 0.2770 & 0.5666 & 0.4551 & 0.4645 & 0.2292 \\
		\midrule
		\textbf{DCGL}      & \textbf{0.2605} & \textbf{0.1119} & \textbf{0.5804} & \textbf{0.2798} & \textbf{0.5727} & \textbf{0.4616} & \textbf{0.4663} & \textbf{0.2317} \\
		\bottomrule
	\end{tabular}
\end{table}

\subsection{Performance Comparison}
Quantitative results in Table~\ref{tab:performance} show that DCGL consistently outperforms all baselines across the four datasets (statistically significant with $p<0.05$). A key observation is DCGL’s exceptional robustness under sparse conditions. On low‑density datasets such as Book‑Crossing (0.19\%) and Amazon‑Book (0.27\%), traditional GNN‑based models (e.g., KGAT, KGCL) struggle because they heavily rely on explicit structural connectivity, which is scarce in such environments. DCGL effectively circumvents this “sparsity trap” by leveraging its decoupled LLM semantic channel. By treating LLM‑derived semantics as an independent view, DCGL constructs implicit semantic bridges that compensate for missing structural links, enabling preference inference based on content understanding even when behavioural paths are broken. This leads to substantial gains, notably surpassing the strongest baseline on Amazon‑Book by a clear margin.

Moreover, DCGL exhibits superior adaptability compared with state‑of‑the‑art LLM‑enhanced baselines such as CoLaKG and CIKGRec. These methods are fundamentally constrained by static fusion strategies that blend semantic and ID signals indiscriminately, often introducing semantic noise for popular items or failing to support cold‑start ones. DCGL resolves this dichotomy through its frequency‑aware gating mechanism, which acts as a dynamic arbiter between the two channels. As evidenced by its consistent lead on both dense (MovieLens) and sparse (DBbook) datasets, DCGL successfully prioritises the LLM channel to alleviate cold‑start issues for tail items while shifting focus to the ID channel for head items. This dynamic calibration effectively balances semantic generalisation with behavioural specificity, validating the necessity of a dual‑channel architecture for handling frequency heterogeneity.

\subsection{Ablation Study}
To evaluate the contribution of each component in DCGL, we conducted ablation studies using fixed hyperparameters, with results on all four core datasets presented in Table~\ref{tab:ablation}. The key findings are as follows: First, removing the LLM semantic channel (\textit{w/o LLM}) leads to substantial performance drops across all datasets, underscoring the critical role of LLM-generated semantic representations in mitigating data sparsity and improving recommendation accuracy. Similarly, discarding the ID information channel (\textit{w/o ID}) also results in performance degradation, indicating that interaction behavior signals captured by the ID channel complement those from the LLM channel to jointly enhance recommendation effectiveness. Replacing the dual‑channel architecture with a simple concatenation of LLM and ID embeddings at the input (\textit{CAT}) yields inferior results, demonstrating the necessity of separate channels for adaptive fusion. Likewise, replacing the frequency‑aware gating with a fixed‑weight scheme (\textit{w/o FREQ}) leads to clear performance decline, verifying that dynamic gating is essential for balancing semantic and behavioural information according to interaction frequency. Furthermore, removing either intra-view augmentation (\textit{w/o AUG}) or inter-view alignment (\textit{w/o ALIGN}) significantly reduces model performance, demonstrating that intra-view augmentation enhances representation robustness while inter-view alignment effectively bridges the heterogeneity between dual-channel representation spaces.

\begin{figure}[t]
	\centering
	\begin{subfigure}{0.23\textwidth}
		\centering
		\includegraphics[width=\linewidth]{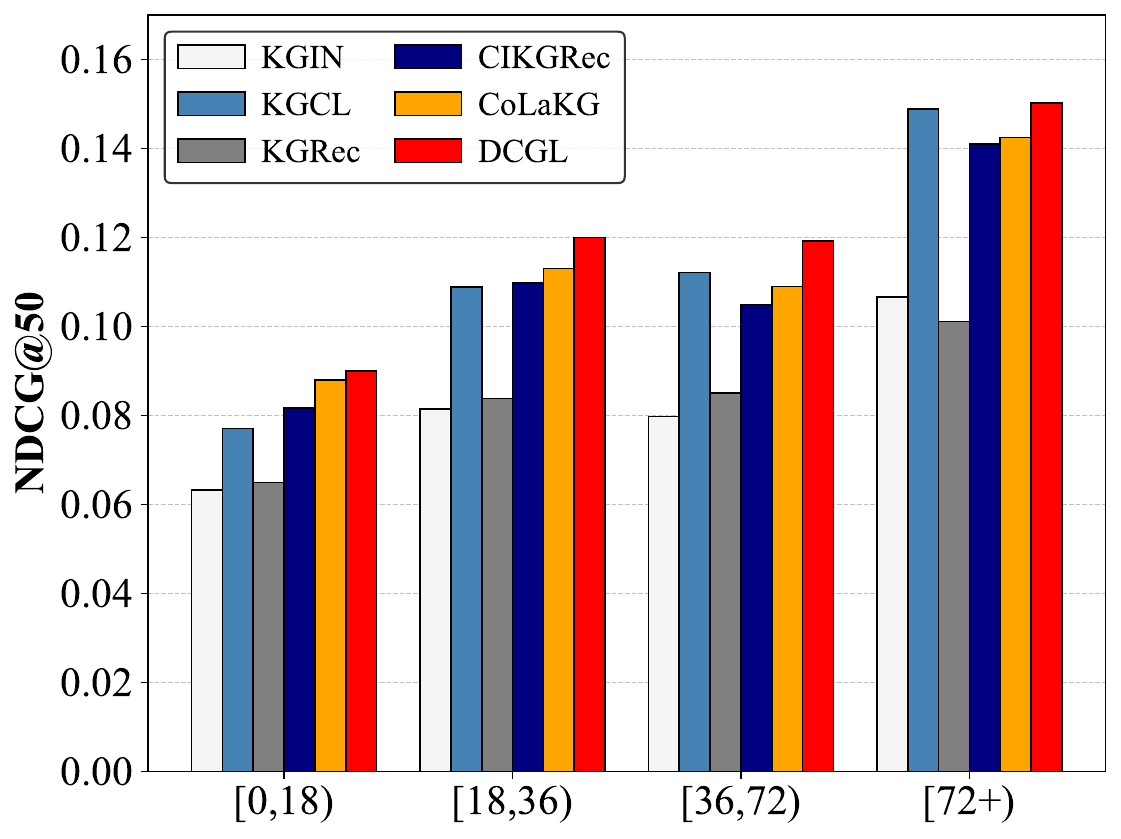}
		\caption{User Group}
		\label{fig:User Group}
	\end{subfigure}
	\hfill
	\begin{subfigure}{0.23\textwidth}
		\centering
		\includegraphics[width=\linewidth]{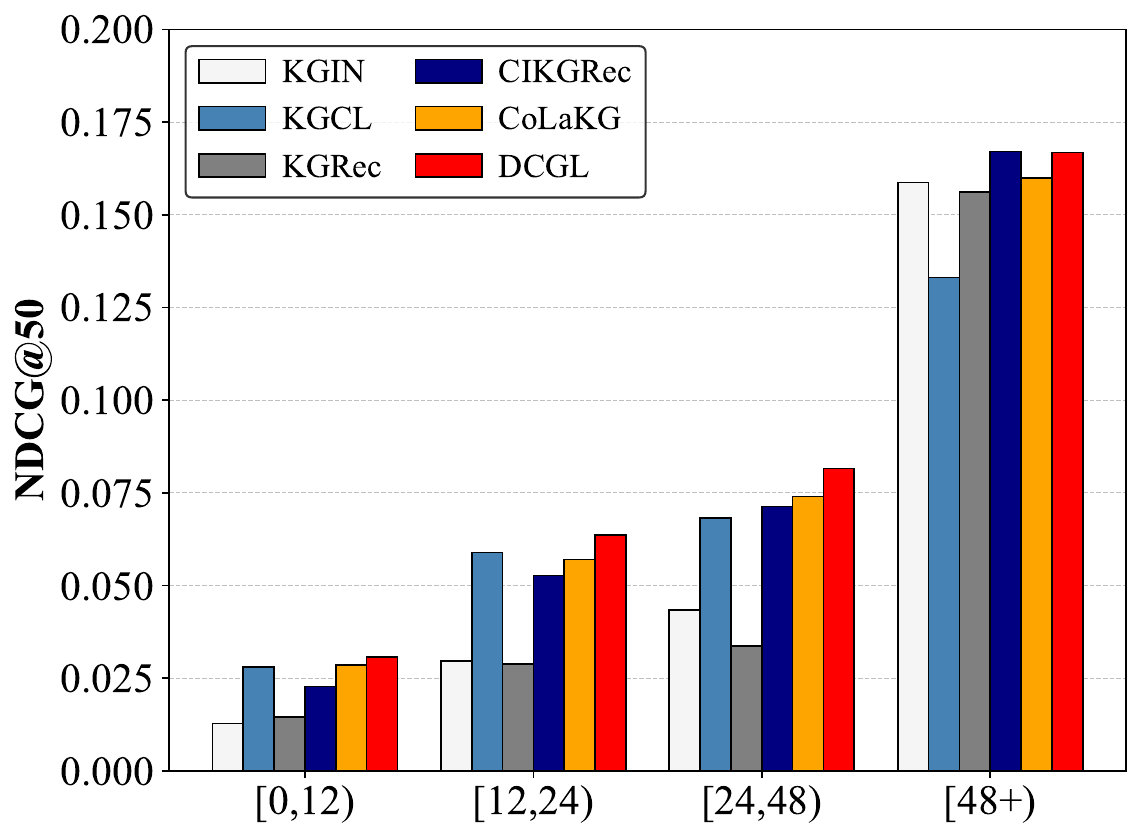}
		\caption{Item Group}
		\label{fig:Item Group}
	\end{subfigure}
	\caption{Performance of DCGL \emph{vs.} Baselines under different user \& item groups (Book-Crossing Dataset).}
	\label{fig:ndcg_comparison}
\end{figure}

\subsection{Group Analysis}
To evaluate the robustness of DCGL across varying data densities, we conduct a group-wise analysis on the Book-Crossing dataset. Users are divided into four intervals based on interaction counts: $[0,18)$, $[18,36)$, $[36,72)$, and $[72+)$, while items are grouped by interaction frequency: $[0,12)$, $[12,24)$, $[24,48)$, and $[48+)$. Figure~\ref{fig:ndcg_comparison} illustrates the performance comparison against competitive baselines. The results indicate that DCGL consistently outperforms alternative methods across all user and item segments, validating its adaptability to both data-scarce and data-rich environments.

\begin{itemize}
	\item \textbf{Sparse Scenarios:} In low-interaction regimes, specifically the user group $[0,18)$ and item group $[0,12)$, DCGL achieves substantial gains over traditional GNNs like KGIN and KGRec, while maintaining a clear lead over other LLM-enhanced baselines. This confirms that the dual-channel architecture effectively harnesses LLM-derived knowledge to mitigate cold-start challenges, substituting missing behavioral signals with rich implicit semantics.
	
	\item \textbf{Dense Scenarios:} In high-interaction regimes (e.g.,  $[72+)$ users,  $[48+)$ items), DCGL maintains superior performance where other methods often exhibit volatility. Unlike purely semantic-enhanced methods that risk introducing noise, DCGL employs the frequency-aware gating mechanism to adaptively prioritize the ID channel ($\uparrow g$). This strategy effectively filters out generic semantic redundancy while fully exploiting the precise collaborative patterns captured by the ID embeddings, ensuring that the rich historical data of active users is leveraged without dilution.
\end{itemize}

In summary, DCGL effectively bridges the information gap for long-tail items while maintaining accuracy for popular ones, validating its adaptive fusion mechanism in balancing semantic generalization and behavioral specificity.

\begin{figure}[t]
	\centering
	\includegraphics[width=\linewidth]{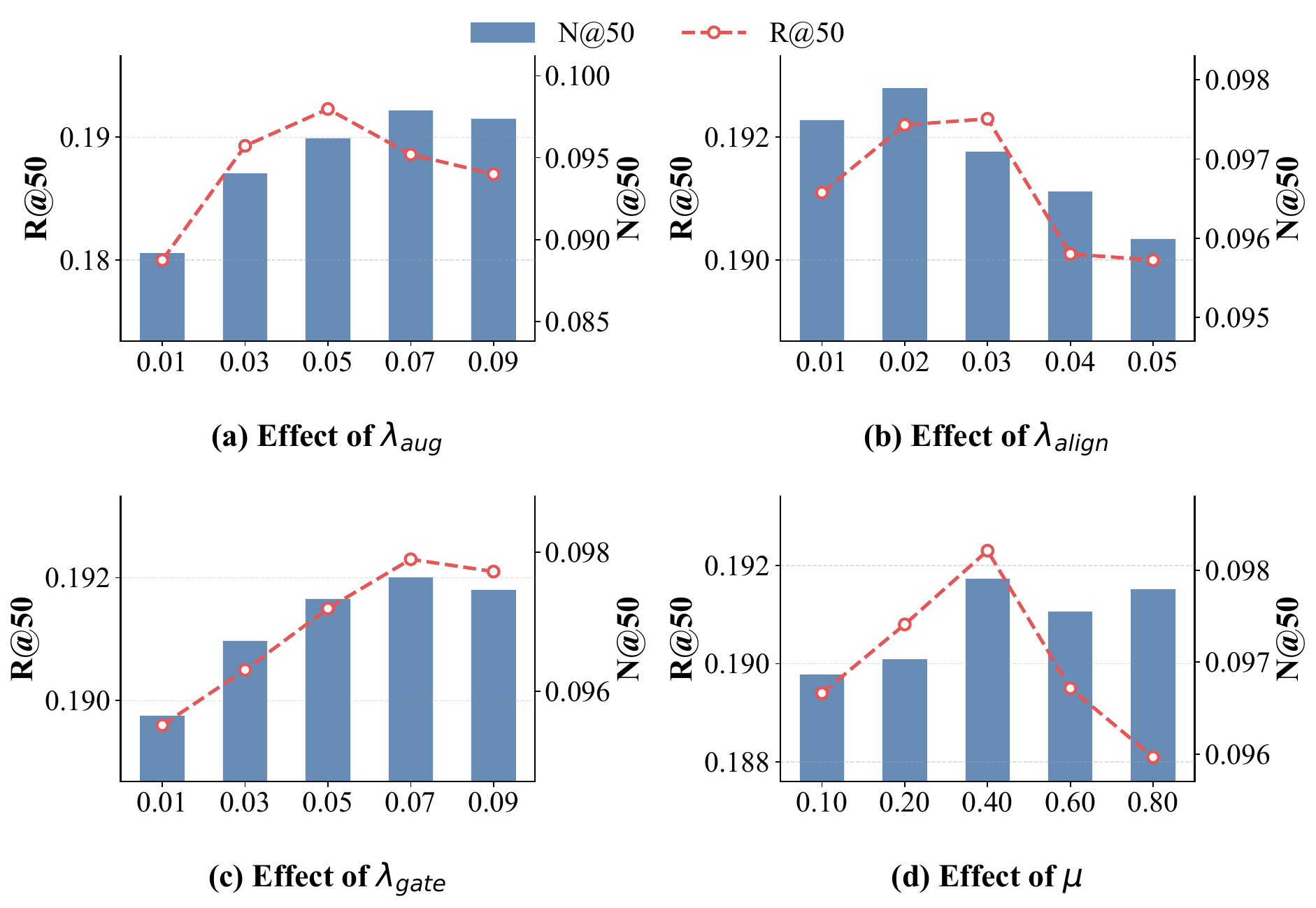} 
	\caption{Hyper-parameter analysis on Book-Crossing.}
	\label{fig:all_paras}
\end{figure}

\subsection{Hyperparameter Sensitivity}
We study DCGL's sensitivity to four key hyperparameters on Book-Crossing: the intra-view contrastive weight $\lambda_{\text{aug}}$, the inter-view contrastive weight $\lambda_{\text{align}}$, the gating regularization weight $\lambda_{\text{gate}}$, and the stability-aware dropout coefficient $\mu$. As illustrated in Figure~\ref{fig:all_paras}(a) and (b), both contrastive weights exhibit a similar trend:  performance first improves and then declines. Small values fail to provide sufficient self-supervised regularization to bridge the semantic gap or enhance robustness. Conversely, excessively large values risk overshadowing the main recommendation objective, leading to suboptimal rankings. A moderate range yields the best trade-off between representation distinctiveness and semantic alignment. Figure~\ref{fig:all_paras}(c) indicates that $\lambda_{\text{gate}}$ is crucial for preventing the gating mechanism from collapsing to trivial solutions (i.e., saturation at 0 or 1), thereby preserving its frequency-aware adaptivity. For the stability coefficient $\mu$ in Figure~\ref{fig:all_paras}(d), an intermediate value achieves the most effective perturbation, confirming that stability-guided pruning effectively removes noise while preserving structural information, outperforming both random and aggressive dropout.

\subsection{Visualization of Gating Interpretability}
\label{sec:gating_vis}

\begin{figure}[t]
	\centering
	\captionsetup[subfigure]{font=footnotesize}
	\begin{subfigure}{0.49\linewidth}
		\centering
		\includegraphics[width=\linewidth]{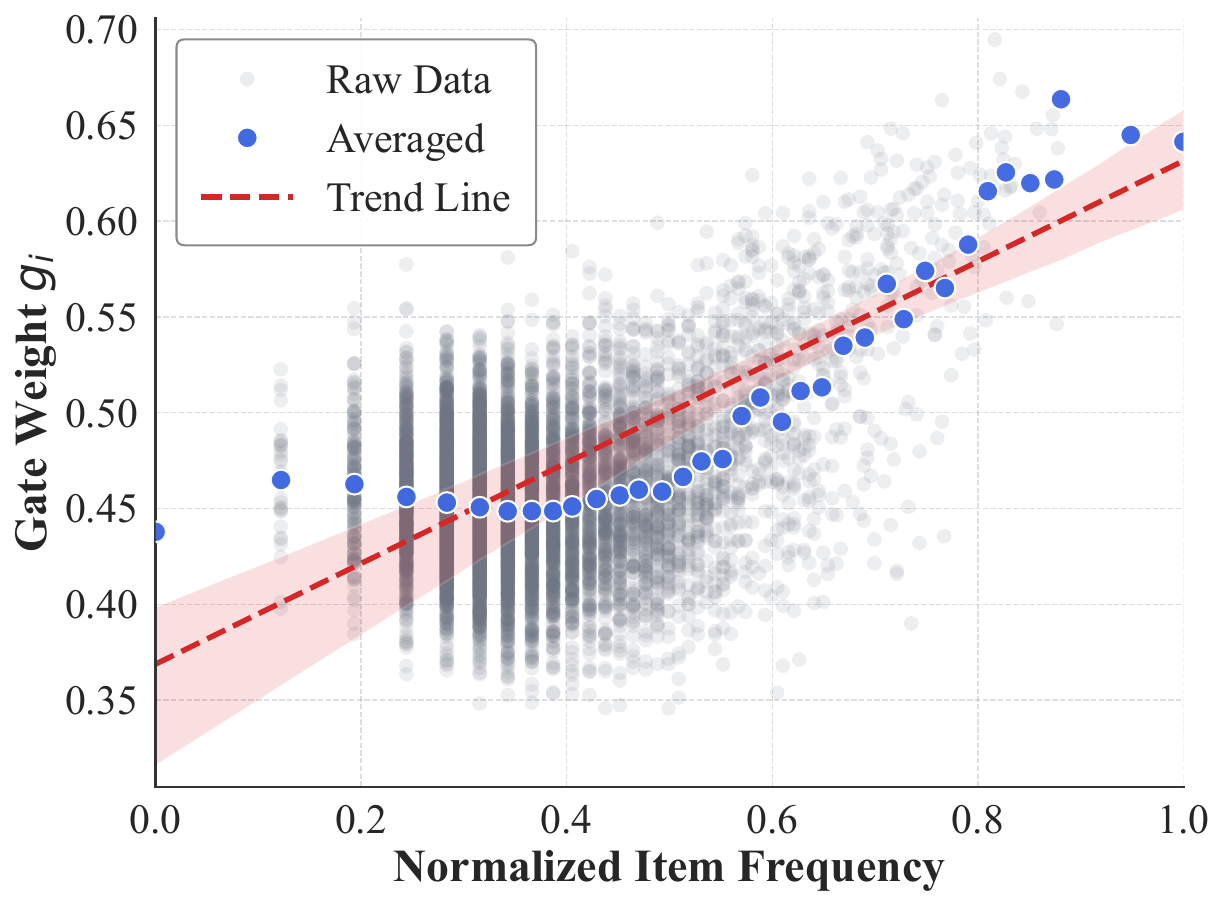}
		\caption{Item popularity vs. Gate weight $g_i$}
		\label{fig:item-gate}
	\end{subfigure}
	\begin{subfigure}{0.49\linewidth}
		\centering
		\includegraphics[width=\linewidth]{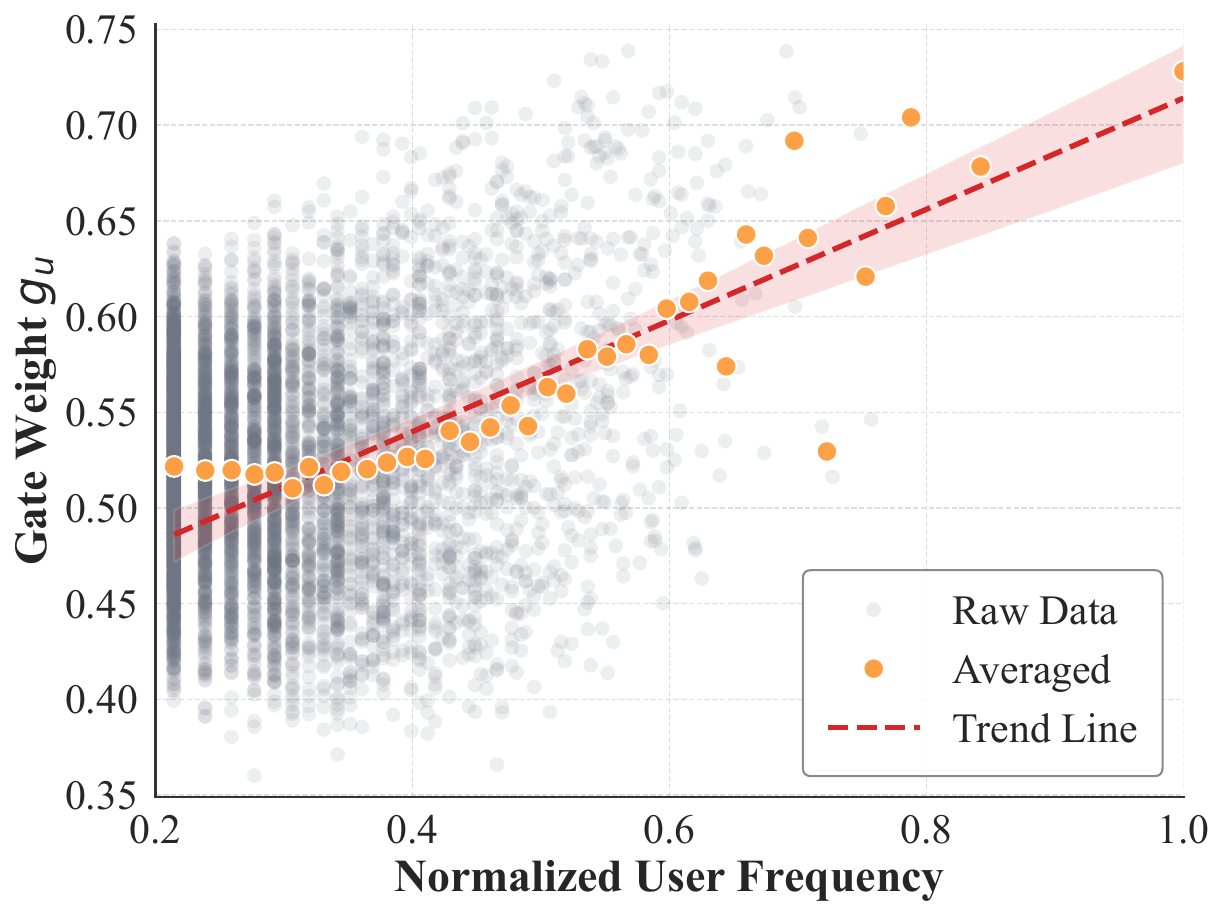}
		\caption{User activity vs. Gate weight $g_u$}
		\label{fig:user-gate}
	\end{subfigure}
	\caption{Visualization of the learned gating weights $g$ w.r.t interaction frequency on Book-Crossing.}
	\label{fig:gate-frequency}
\end{figure}

To verify whether DCGL effectively resolves the frequency heterogeneity challenge highlighted in our Motivating Example, we visualize the learned gating weights on the Book-Crossing dataset. Figure~\ref{fig:gate-frequency} illustrates the relationship between normalized interaction frequency and the learned gate weight $g$ (which controls the contribution of the ID channel). As shown in Figure~\ref{fig:gate-frequency}, the clear positive correlation between interaction frequency and gate weight $g$ provides empirical evidence that our model successfully adapts its fusion strategy across the sparsity spectrum:

\begin{itemize}
	\item \textbf{Response to the ``Bob'' Scenario (Sparse Regime):} In the low-frequency region (left side of the plots), the model consistently learns lower gate weights (lower $g$). This aligns perfectly with the needs of cold-start users like Bob. Since their behavioral signals are weak and noise-prone, DCGL automatically suppresses the unreliable ID channel and assigns a larger weight $(1-g)$ to the LLM channel. This mechanism effectively leverages clear semantic cues to compensate for the lack of interaction data, preventing the ``noise overwhelming'' issue described in the motivation.
	
	\item \textbf{Response to the ``Alice'' Scenario (Dense Regime):} In the high-frequency region (right side of the plots), the gate weights significantly increase, indicating a dominance of the ID channel. This mirrors the scenario of active users like Alice. Given her strong historical patterns, the model prioritizes the precise collaborative signals captured by the ID embeddings. Crucially, by limiting the weight of the LLM channel in this regime, DCGL avoids the ``blurred representations'' caused by indiscriminate semantic infusion, ensuring that Alice's precise personal preferences are not diluted by generic semantic information.
\end{itemize}

This visualization confirms that DCGL breaks the vicious cycle of static fusion. Instead of a uniform view, it achieves a context-sensitive balance: relying on semantics for grounding in sparse cases while preserving behavioral precision in dense cases.

\section{Conclusion}
\label{sec:conclusion}

In this paper, we proposed DCGL, a novel dual-channel graph learning framework designed to break the cascading limitation in knowledge-aware recommendation. Unlike conventional methods that suffer from signal interference due to static fusion, DCGL structurally decouples frequency-agnostic LLM semantics from frequency-sensitive behavioral signals. Through a synergistic combination of multi-level contrastive learning and frequency-aware gated fusion, our approach effectively resolves the trade-off between semantic generalization and behavioral specificity. Extensive experiments on four real-world datasets demonstrate that DCGL not only consistently outperforms state-of-the-art baselines but also achieves a context-sensitive balance: leveraging external knowledge to alleviate sparsity for cold-start entities while preserving precise collaborative patterns for active ones. Furthermore, since LLM-based semantic extraction is performed offline, DCGL requires no real-time LLM inference, ensuring high efficiency for practical deployment.

\bibliographystyle{ACM-Reference-Format}
\balance
\bibliography{sigir2026}

@String{Computing = "Computing" }

@String{Springer = "Springer-Verlag" }

@ArtifactSoftware{R,
    title = {R: A Language and Environment for Statistical Computing},
    author = {{R Core Team}},
    organization = {R Foundation for Statistical Computing},
    address = {Vienna, Austria},
    year = {2019},
    url = {https://www.R-project.org/},
}

@inproceedings{zhang2016collaborative,
  title={Collaborative knowledge base embedding for recommender systems},
  author={Zhang, Fuzheng and Yuan, Nicholas Jing and Lian, Defu and Xie, Xing and Ma, Wei-Ying},
  booktitle={Proceedings of the 22nd ACM SIGKDD international conference on knowledge discovery and data mining},
  pages={353--362},
  year={2016}
}

@inproceedings{yu2014personalized,
	title={Personalized entity recommendation: A heterogeneous information network approach},
	author={Yu, Xiao and Ren, Xiang and Sun, Yizhou and Gu, Quanquan and Sturt, Bradley and Khandelwal, Urvashi and Norick, Brandon and Han, Jiawei},
	booktitle={Proceedings of the 7th ACM international conference on Web search and data mining},
	pages={283--292},
	year={2014}
}

@inproceedings{yang2022knowledge,
	title={Knowledge graph contrastive learning for recommendation},
	author={Yang, Yuhao and Huang, Chao and Xia, Lianghao and Li, Chenliang},
	booktitle={Proceedings of the 45th international ACM SIGIR conference on research and development in information retrieval},
	pages={1434--1443},
	year={2022}
}

@inproceedings{wang2019kgat,
	title={Kgat: Knowledge graph attention network for recommendation},
	author={Wang, Xiang and He, Xiangnan and Cao, Yixin and Liu, Meng and Chua, Tat-Seng},
	booktitle={Proceedings of the 25th ACM SIGKDD international conference on knowledge discovery \& data mining},
	pages={950--958},
	year={2019}
}

@inproceedings{yang2023knowledge,
	title={Knowledge graph self-supervised rationalization for recommendation},
	author={Yang, Yuhao and Huang, Chao and Xia, Lianghao and Huang, Chunzhen},
	booktitle={Proceedings of the 29th ACM SIGKDD conference on knowledge discovery and data mining},
	pages={3046--3056},
	year={2023}
}

@inproceedings{hu2025bridging,
	title={Bridging the user-side knowledge gap in knowledge-aware recommendations with large language models},
	author={Hu, Zheng and Li, Zhe and Jiao, Ziyun and Nakagawa, Satoshi and Deng, Jiawen and Cai, Shimin and Zhou, Tao and Ren, Fuji},
	booktitle={Proceedings of the AAAI Conference on Artificial Intelligence},
	volume={39},
	number={11},
	pages={11799--11807},
	year={2025}
}

@article{zhao2024recommender,
	title={Recommender systems in the era of large language models (llms)},
	author={Zhao, Zihuai and Fan, Wenqi and Li, Jiatong and Liu, Yunqing and Mei, Xiaowei and Wang, Yiqi and Wen, Zhen and Wang, Fei and Zhao, Xiangyu and Tang, Jiliang and others},
	journal={IEEE Transactions on Knowledge and Data Engineering},
	volume={36},
	number={11},
	pages={6889--6907},
	year={2024},
	publisher={IEEE}
}

@inproceedings{li2024large,
	title={Large language models for generative recommendation: A survey and visionary discussions},
	author={Li, Lei and Zhang, Yongfeng and Liu, Dugang and Chen, Li},
	booktitle={Proceedings of the 2024 Joint International Conference on Computational Linguistics, Language Resources and Evaluation (LREC-COLING 2024)},
	pages={10146--10159},
	year={2024}
}

@article{chen2024large,
	title={When large language models meet personalization: Perspectives of challenges and opportunities},
	author={Chen, Jin and Liu, Zheng and Huang, Xu and Wu, Chenwang and Liu, Qi and Jiang, Gangwei and Pu, Yuanhao and Lei, Yuxuan and Chen, Xiaolong and Wang, Xingmei and others},
	journal={World Wide Web},
	volume={27},
	number={4},
	pages={42},
	year={2024},
	publisher={Springer}
}

@article{wu2024survey,
	title={A survey on large language models for recommendation},
	author={Wu, Likang and Zheng, Zhi and Qiu, Zhaopeng and Wang, Hao and Gu, Hongchao and Shen, Tingjia and Qin, Chuan and Zhu, Chen and Zhu, Hengshu and Liu, Qi and others},
	journal={World Wide Web},
	volume={27},
	number={5},
	pages={60},
	year={2024},
	publisher={Springer}
}

@inproceedings{hou2024large,
	title={Large language models are zero-shot rankers for recommender systems},
	author={Hou, Yupeng and Zhang, Junjie and Lin, Zihan and Lu, Hongyu and Xie, Ruobing and McAuley, Julian and Zhao, Wayne Xin},
	booktitle={European Conference on Information Retrieval},
	pages={364--381},
	year={2024},
	organization={Springer}
}

@inproceedings{bao2023tallrec,
	title={Tallrec: An effective and efficient tuning framework to align large language model with recommendation},
	author={Bao, Keqin and Zhang, Jizhi and Zhang, Yang and Wang, Wenjie and Feng, Fuli and He, Xiangnan},
	booktitle={Proceedings of the 17th ACM conference on recommender systems},
	pages={1007--1014},
	year={2023}
}

@inproceedings{gao2025sprec,
	title={Sprec: Self-play to debias llm-based recommendation},
	author={Gao, Chongming and Chen, Ruijun and Yuan, Shuai and Huang, Kexin and Yu, Yuanqing and He, Xiangnan},
	booktitle={Proceedings of the ACM on Web Conference 2025},
	pages={5075--5084},
	year={2025}
}

@inproceedings{li2025g,
	title={G-refer: Graph retrieval-augmented large language model for explainable recommendation},
	author={Li, Yuhan and Zhang, Xinni and Luo, Linhao and Chang, Heng and Ren, Yuxiang and King, Irwin and Li, Jia},
	booktitle={Proceedings of the ACM on Web Conference 2025},
	pages={240--251},
	year={2025}
}

@inproceedings{wei2024llmrec,
	title={Llmrec: Large language models with graph augmentation for recommendation},
	author={Wei, Wei and Ren, Xubin and Tang, Jiabin and Wang, Qinyong and Su, Lixin and Cheng, Suqi and Wang, Junfeng and Yin, Dawei and Huang, Chao},
	booktitle={Proceedings of the 17th ACM international conference on web search and data mining},
	pages={806--815},
	year={2024}
}

@inproceedings{ren2024representation,
	title={Representation learning with large language models for recommendation},
	author={Ren, Xubin and Wei, Wei and Xia, Lianghao and Su, Lixin and Cheng, Suqi and Wang, Junfeng and Yin, Dawei and Huang, Chao},
	booktitle={Proceedings of the ACM web conference 2024},
	pages={3464--3475},
	year={2024}
}

@inproceedings{jiang2025reclm,
	title = "{R}ec{LM}: Recommendation Instruction Tuning",
	author = "Jiang, Yangqin  and
	Yang, Yuhao  and
	Xia, Lianghao  and
	Luo, Da  and
	Lin, Kangyi  and
	Huang, Chao",
	editor = "Che, Wanxiang  and
	Nabende, Joyce  and
	Shutova, Ekaterina  and
	Pilehvar, Mohammad Taher",
	booktitle = "Proceedings of the 63rd Annual Meeting of the Association for Computational Linguistics (Volume 1: Long Papers)",
	month = jul,
	year = "2025",
	publisher = "Association for Computational Linguistics",
	url = "https://aclanthology.org/2025.acl-long.751/",
	pages = "15443--15459"
}

@inproceedings{cui2025comprehending,
	title={Comprehending knowledge graphs with large language models for recommender systems},
	author={Cui, Ziqiang and Weng, Yunpeng and Tang, Xing and Lyu, Fuyuan and Liu, Dugang and He, Xiuqiang and Ma, Chen},
	booktitle={Proceedings of the 48th International ACM SIGIR Conference on Research and Development in Information Retrieval},
	pages={1229--1239},
	year={2025}
}

@article{gao2023survey,
	title={A survey of graph neural networks for recommender systems: Challenges, methods, and directions},
	author={Gao, Chen and Zheng, Yu and Li, Nian and Li, Yinfeng and Qin, Yingrong and Piao, Jinghua and Quan, Yuhan and Chang, Jianxin and Jin, Depeng and He, Xiangnan and others},
	journal={ACM Transactions on Recommender Systems},
	volume={1},
	number={1},
	pages={1--51},
	year={2023},
	publisher={ACM New York, NY, USA}
}

@inproceedings{he2017neural,
	title={Neural factorization machines for sparse predictive analytics},
	author={He, Xiangnan and Chua, Tat-Seng},
	booktitle={Proceedings of the 40th International ACM SIGIR conference on Research and Development in Information Retrieval},
	pages={355--364},
	year={2017}
}

@inproceedings{he2020lightgcn,
	title={Lightgcn: Simplifying and powering graph convolution network for recommendation},
	author={He, Xiangnan and Deng, Kuan and Wang, Xiang and Li, Yan and Zhang, Yongdong and Wang, Meng},
	booktitle={Proceedings of the 43rd International ACM SIGIR conference on research and development in Information Retrieval},
	pages={639--648},
	year={2020}
}

@inproceedings{togashi2021alleviating,
	title={Alleviating cold-start problems in recommendation through pseudo-labelling over knowledge graph},
	author={Togashi, Riku and Otani, Mayu and Satoh, Shin'ichi},
	booktitle={Proceedings of the 14th ACM international conference on web search and data mining},
	pages={931--939},
	year={2021}
}

@inproceedings{wang2021learning,
	title={Learning intents behind interactions with knowledge graph for recommendation},
	author={Wang, Xiang and Huang, Tinglin and Wang, Dingxian and Yuan, Yancheng and Liu, Zhenguang and He, Xiangnan and Chua, Tat-Seng},
	booktitle={Proceedings of the web conference 2021},
	pages={878--887},
	year={2021}
}

@inproceedings{zou2022multi,
	title={Multi-level cross-view contrastive learning for knowledge-aware recommender system},
	author={Zou, Ding and Wei, Wei and Mao, Xian-Ling and Wang, Ziyang and Qiu, Minghui and Zhu, Feida and Cao, Xin},
	booktitle={Proceedings of the 45th international ACM SIGIR conference on research and development in information retrieval},
	pages={1358--1368},
	year={2022}
}

@inproceedings{cao2019unifying,
	title={Unifying knowledge graph learning and recommendation: Towards a better understanding of user preferences},
	author={Cao, Yixin and Wang, Xiang and He, Xiangnan and Hu, Zikun and Chua, Tat-Seng},
	booktitle={The world wide web conference},
	pages={151--161},
	year={2019}
}

@inproceedings{dong2017hybrid,
	title={A hybrid collaborative filtering model with deep structure for recommender systems},
	author={Dong, Xin and Yu, Lei and Wu, Zhonghuo and Sun, Yuxia and Yuan, Lingfeng and Zhang, Fangxi},
	booktitle={Proceedings of the AAAI Conference on artificial intelligence},
	volume={31},
	number={1},
	year={2017}
}

@article{noia2016sprank,
	title={Sprank: Semantic path-based ranking for top-n recommendations using linked open data},
	author={Noia, Tommaso Di and Ostuni, Vito Claudio and Tomeo, Paolo and Sciascio, Eugenio Di},
	journal={ACM Transactions on Intelligent Systems and Technology (TIST)},
	volume={8},
	number={1},
	pages={1--34},
	year={2016},
	publisher={ACM New York, NY, USA}
}

@inproceedings{he2016ups,
	title={Ups and downs: Modeling the visual evolution of fashion trends with one-class collaborative filtering},
	author={He, Ruining and McAuley, Julian},
	booktitle={proceedings of the 25th international conference on world wide web},
	pages={507--517},
	year={2016}
}

@inproceedings{xie2022contrastive,
	title={Contrastive learning for sequential recommendation},
	author={Xie, Xu and Sun, Fei and Liu, Zhaoyang and Wu, Shiwen and Gao, Jinyang and Zhang, Jiandong and Ding, Bolin and Cui, Bin},
	booktitle={2022 IEEE 38th international conference on data engineering (ICDE)},
	pages={1259--1273},
	year={2022},
	organization={IEEE}
}

@inproceedings{wu2021self,
  title={Self-supervised graph learning for recommendation},
  author={Wu, Jiancan and Wang, Xiang and Feng, Fuli and He, Xiangnan and Chen, Liang and Lian, Jianxun and Xie, Xing},
  booktitle={Proceedings of the 44th international ACM SIGIR conference on research and development in information retrieval},
  pages={726--735},
  year={2021}
}

@inproceedings{yu2022graph,
  title={Are graph augmentations necessary? simple graph contrastive learning for recommendation},
  author={Yu, Junliang and Yin, Hongzhi and Xia, Xin and Chen, Tong and Cui, Lizhen and Nguyen, Quoc Viet Hung},
  booktitle={Proceedings of the 45th international ACM SIGIR conference on research and development in information retrieval},
  pages={1294--1303},
  year={2022}
}

@article{ai2018learning,
  title={Learning heterogeneous knowledge base embeddings for explainable recommendation},
  author={Ai, Qingyao and Azizi, Vahid and Chen, Xu and Zhang, Yongfeng},
  journal={Algorithms},
  volume={11},
  number={9},
  pages={137},
  year={2018},
  publisher={MDPI}
}

@inproceedings{wang2018dkn,
	title={DKN: Deep knowledge-aware network for news recommendation},
	author={Wang, Hongwei and Zhang, Fuzheng and Xie, Xing and Guo, Minyi},
	booktitle={Proceedings of the 2018 world wide web conference},
	pages={1835--1844},
	year={2018}
}

@article{Wang_Wang_Xu_He_Cao_Chua_2019, title={Explainable Reasoning over Knowledge Graphs for Recommendation}, volume={33}, url={https://ojs.aaai.org/index.php/AAAI/article/view/4470}, DOI={10.1609/aaai.v33i01.33015329}, number={01}, journal={Proceedings of the AAAI Conference on Artificial Intelligence}, author={Wang, Xiang and Wang, Dingxian and Xu, Canran and He, Xiangnan and Cao, Yixin and Chua, Tat-Seng}, year={2019}, month={Jul.}, pages={5329-5336} }

@inproceedings{zou2024knowledge,
	title={Knowledge enhanced multi-intent transformer network for recommendation},
	author={Zou, Ding and Wei, Wei and Zhu, Feida and Xu, Chuanyu and Zhang, Tao and Huo, Chengfu},
	booktitle={Companion proceedings of the ACM web conference 2024},
	pages={1--9},
	year={2024}
}

@article{bordes2013translating,
	title={Translating embeddings for modeling multi-relational data},
	author={Bordes, Antoine and Usunier, Nicolas and Garcia-Duran, Alberto and Weston, Jason and Yakhnenko, Oksana},
	journal={Advances in neural information processing systems},
	volume={26},
	year={2013}
}

@article{rendle2012bpr,
	title={BPR: Bayesian personalized ranking from implicit feedback},
	author={Rendle, Steffen and Freudenthaler, Christoph and Gantner, Zeno and Schmidt-Thieme, Lars},
	journal={arXiv preprint arXiv:1205.2618},
	year={2012}
}

@inproceedings{li2025hypercomplex,
	title={Hypercomplex knowledge graph-aware recommendation},
	author={Li, Anchen and Yang, Bo and Huo, Huan and Hussain, Farookh and Xu, Guandong},
	booktitle={Proceedings of the 48th international ACM SIGIR conference on research and development in information retrieval},
	pages={2017--2026},
	year={2025}
}

@inproceedings{li2025lightkg,
	title={LightKG: Efficient Knowledge-Aware Recommendations with Simplified GNN Architecture},
	author={Li, Yanhui and Wang, Dongxia and Sun, Zhu and Zhang, Haonan and Guo, Huizhong},
	booktitle={Proceedings of the 31st ACM SIGKDD Conference on Knowledge Discovery and Data Mining V. 2},
	pages={1577--1588},
	year={2025}
}

@inproceedings{zhao2024breaking,
	title={Breaking the barrier: utilizing large language models for industrial recommendation systems through an inferential knowledge graph},
	author={Zhao, Qian and Qian, Hao and Liu, Ziqi and Zhang, Gong-Duo and Gu, Lihong},
	booktitle={Proceedings of the 33rd ACM International Conference on Information and Knowledge Management},
	pages={5086--5093},
	year={2024}
}

\end{document}